\documentclass[transmag]{IEEEtran}
\usepackage{latexsym}
\usepackage{graphicx}
\usepackage{amsfonts,amssymb,amsmath}
\usepackage{hyperref}

\usepackage{cite}
\usepackage{nameref}
\usepackage[tight,footnotesize]{subfigure}
\usepackage{color}
\usepackage{multirow}
\usepackage{longtable}
\usepackage{float}
\usepackage[section]{placeins}

\def\BibTeX{{\rm B\kern-.05em{\sc i\kern-.025em b}\kern-.08em T\kern-.1667em\lower.7ex\hbox{E}\kern-.125emX}}
\begin{document}

\title{High Utilization Energy-Aware Real-Time Inference Deep Convolutional Neural Network Accelerator}

\author{Kuan-Ting Lin, Ching-Te Chiu, Jheng-Yi Chang, Shi-Zong Huang, and Yu-Ting Li
\thanks{
}
\thanks{Kuan-Ting Lin is with the Department of Computer Science, National Tsing Hua University, Hsinchu, Taiwan (e-mail: jerry8404@gapp.nthu.edu.tw).}
\thanks{Ching-Te Chiu is with the Department of Computer Science, National Tsing Hua University, Hsinchu, Taiwan (e-mail: ctchiu@cs.nthu.edu.tw).}
\thanks{Jheng-Yi Chang is with 
the Institute of Communications Engineering, National Tsing Hua University, Hsinchu, Taiwan (e-mail: a2282094710@gmail.com.}
\thanks{Shi-Zong Huang is with 
the Institute of Communications Engineering, National Tsing Hua University, Hsinchu, Taiwan (e-mail: alex605103@gmail.com).}
\thanks{Yu-Ting Li is with the Department of Computer Science, National Tsing Hua University, Hsinchu, Taiwan (e-mail: yutingli0509@gmail.com).}}

\IEEEtitleabstractindextext{\begin{abstract}Deep convolution Neural Network (DCNN) has been widely used in computer vision tasks. However, for edge devices even inference has too large computational complexity and data access amount. The inference latency of state-of-the-art models are impractical for real-world applications. In this paper, we propose a high utilization energy-aware real-time inference deep convolutional neural network accelerator, which improves the performance of the current accelerators. First, we use the 1x1 size convolution kernel as the smallest unit of the computing unit. Then we design suitable computing unit based on the requirements of each model. Secondly, we use Reuse Feature SRAM to store the output of the current layer in the chip and use the value as the input of the next layer. Moreover, we import Output Reuse Strategy and Ring Stream Dataflow to reduce the amount of data exchange between chips and DRAM. Finally, we present On-fly Pooling Module to let the calculation of the Pooling layer directly complete in the chip. With the aid of the proposed method, the implemented acceleration chip has an extremely high hardware utilization rate. We reduce a generous amount of data transfer on the specific module, ECNN. Compared to the methods without reuse strategy, we can reduce 533 times of data access amount. At the same time, we have enough computing power to perform real-time execution of the existing image classification model, VGG16 and MobileNet. Compared with the design in VWA, we can speed up 7.52 times and have 1.92x energy efficiency.\end{abstract}

\begin{IEEEkeywords}
CNN, Accelerator, Energy-Aware, Real-Time Inference, High Utilization.
\end{IEEEkeywords}}

\maketitle

\section{INTRODUCTION}

\begin{figure}[h]
    \centering
    \includegraphics[width=8cm]{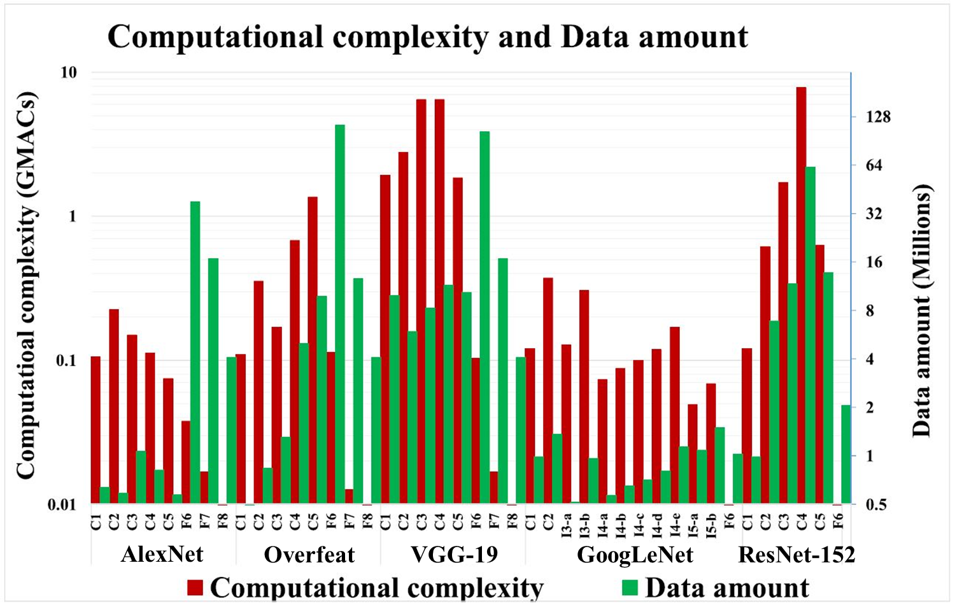}
    \caption{Computational complexity and data amount of the state-of-the-art models.~\cite{lin2017data}.}
    \label{fig:Computation and data}
\end{figure}

\IEEEPARstart{F}{or} the recent years, deep convolution neural networks (DCNN) have been verified to apply on tasks of computer vision, face recognition~\cite{deng2018arcface, huang2020curricularface, liu2017sphereface, sun2013deep}, object detection~\cite{redmon2016you,ren2015faster,liu2016ssd,he2017mask}, action recognition~\cite{I3D,C3D,TwoStream,TSN} and gesture recognition~\cite{wang2020fast,ma2018design,li2015feature,gao2017static}.
Great success of DCNN has brought a lot of convenience to human life with these real-world applications. Although the current graphics processing unit (GPU) can provide high computing performance for training or inference DCNN models. But for the edge devices, even the stage of inference still have two problems to be solved, high computational complexity and amount of external memory access. As shown in Fig.~\ref{fig:Computation and data}, DCNN models needs a lot of computation complexity and data amount to execute the inference stage.

Since the large energy cost of GPU, many research of CNN accelerator design have been conducted~\cite{aimar2018nullhop,chang2019vwa,chen2016eyeriss,du2017reconfigurable,farabet2009cnp,hsu2020essa,huang2019accelerating,huang2019ecnn,chen2014diannao,chen2014dadiannao,luo2016dadiannao}. The first CNN accelerator~\cite{farabet2009cnp} is implement on the FPGA board, but with the limited performance. Then, \cite{zhang2015optimizing} uses loop-tiling technique to optimize the performance and the communication to computation ratio for the accelerator and use the roofline~\cite{williams2009roofline} model for find out the best tiling factors in loop tiling. In the family of DianNao~\cite{chen2014diannao,chen2014dadiannao,luo2016dadiannao}, they use the internal buffer to reduce the of chip data access for decrease the energy consumption. In \cite{huang2019accelerating} use the ping-pong SRAM to reduce the data access amount from external memory. The most important, \cite{mcdanel2019full} shows that if we want to optimize convolution, not only design the hardware accelerator but the model structure. 

Thus, we want to design a CNN hardware accelerator which can do real-time inference in edge device. Therefore, there are four crucial issues in our accelerator design, core area size, on-chip SRAM size, enough throughput for real-time inference state-of-the-art models and power efficiency. For the core area size and the on-chip SRAM size, in order to set our accelerator to embedded system, we design to limit the core area size to 4 $mm^2$, and 300 KB for the on-chip SRAM size. And for the enough throughput for real-time inference state-of-the-art models, we want our accelerator can provide over 600 Gops/s for inference model VGG-16 in real-time. Moreover, we want the energy efficiency of the proposed accelerator is higher than \cite{chang2019vwa}'s approach to make ours accelerator more suitable to the edge devices. 

The main contributions of this thesis are presented below:

1. We propose a high utilization energy-aware real-time inference DCNN accelerator design to tackle three important issues, hardware utilization, external memory access and computation complexity.

2. The hardware implementation of the proposed accelerator architecture under the TSMC 40 nm technology reaches 1.152 Tops/s with 554.57 mW total power in 3.59 $mm^2$ area size.

3. Compared the proposed CNN accelerator to \cite{chang2019vwa} on VGG16, we have higher hardware utilization in convolution layer 1 to speed up the inference time about 7.5x and reduce the data access amount of external DRAM about 1.4x, which is the most energy consumption part for the accelerator inference the CNN model. Therefore, the energy efficiency of the proposed accelerator reach 2.08 Tops/W, which is 1.92x higher than \cite{chang2019vwa}.

The rest of this thesis is organized as follows: Chapter II describes the basic of CNN, some prior works and the embedded CNN structure. Chapter III introduces the proposed accelerator architecture and the proposed methods. Chapter IV the theoretical evaluation and the implementation of proposed accelerator have been provided. Finally, Chapter V presents our conclusions.

\section{Related Work}
\label{sec:related_works}

\subsection{CNN Operations}
\label{s:CNN_operation}
\begin{figure}[h]
    \centering
    \includegraphics[width=8cm]{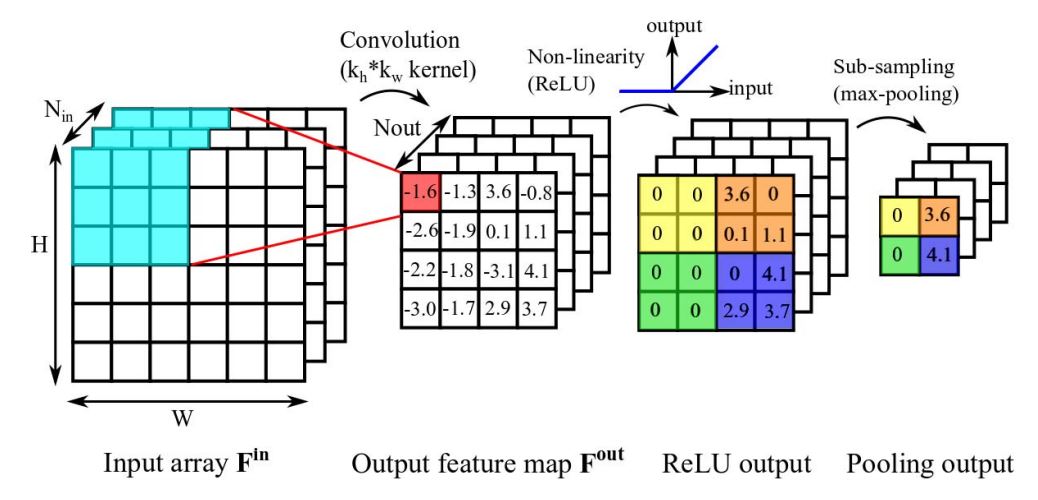}
    \caption{Three stages of CNN operation.}
    \label{fig:CNN}
\end{figure}

CNN model has first been proposed in \cite{lecun1989backpropagation} many years ago. with the aid of powerful and affordable GPUs, a plenty of CNN model design research have been proposed, such as AlexNet~\cite{krizhevsky2012imagenet}, ResNet~\cite{he2016deep}, DenseNet~\cite{huang2017densely} and FaceNet~\cite{schroff2015facenet}. And there are three stages to perform inference, convolution stage, activation stage and pooling stage as shows in Fig.~\ref{fig:CNN}. The convolution stage inputs a image as input feature pixels convolve with kernel pixels in order to obtain output feature maps. After getting output feature map, the CNN operation enters the activation stage. In the activation stage, this operation convert the floating values in the output feature maps into several constant to let the whole model easier to process. Take ReLu activation as an example, when passing ReLU activation, the floating value smaller than zero in the output feature maps are converted into zero. On the contrary, if the floating values are larger than zero, the value remains the same after passing the ReLU activation. Next, the Pooling stage is designed to do the operation of down-sampling. It splits the ReLu output into several pieces then only chooses one value to represent that piece. For example, when operation Max-pooling, it splits the ReLu output with 2x2 blocks then it only keeps the maximum value to represent the corresponding block. Therefore, one block with four pixels is converted into one pixel after the pooling stage.

\subsection{ASIC Based CNN Accelerator}
\label{s:ASIC}
\begin{figure}[h]
    \centering
    \includegraphics[width=8cm]{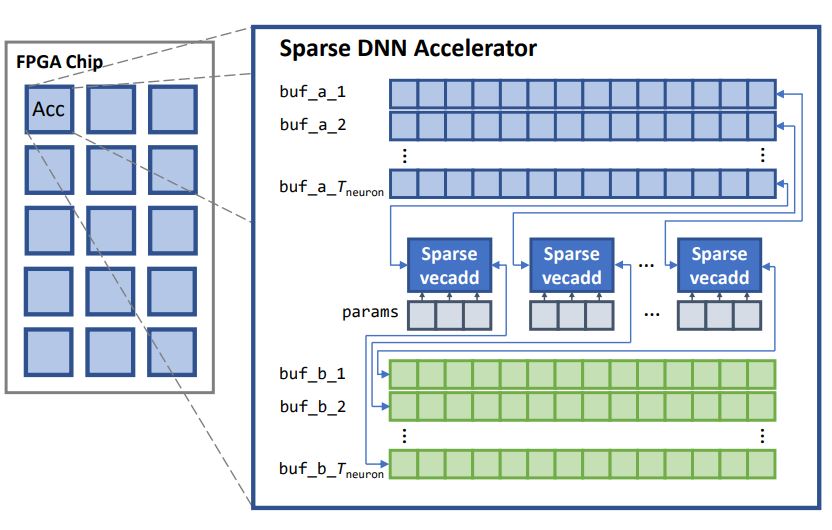}
    \caption{Overview of sparse DNN accelerator architecture.~\cite{huang2019accelerating}}
    \label{fig:sparse_accelerator}
\end{figure}

In \cite{huang2019accelerating}, They use the sparsity of DCNN parameters to reduce data access amount and computational complexity of perform inference. As shown in Fig.~\ref{fig:sparse_accelerator}, they use ping-pong SRAM for store the output feature map in the on-chip SRAM and use for the input feature map of next convolution layer to reduce the data access amount. But it needs a work allocator to allocate works to process element and the CNN models are not always sparsity.

\begin{figure}[h]
    \centering
    \includegraphics[width=8cm]{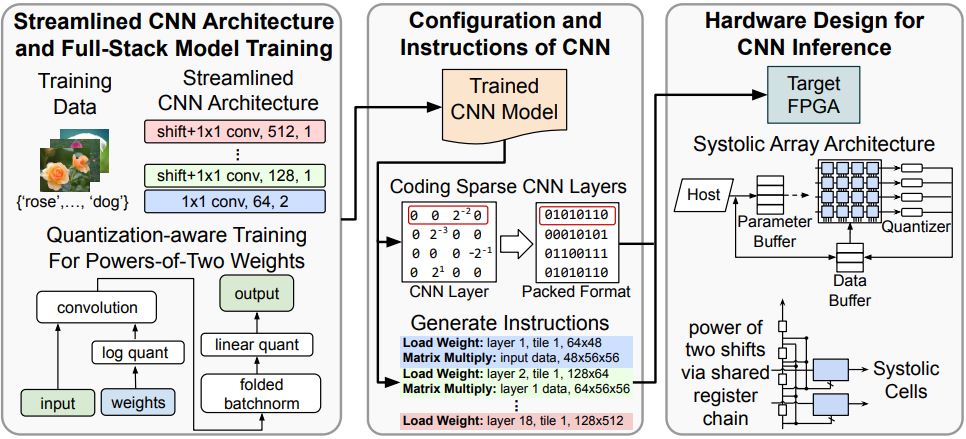}
    \caption{Overview of the full-stack optimization framework.~\cite{mcdanel2019full}}
    \label{fig:full-stack}
\end{figure}

In \cite{mcdanel2019full},they proposed the full-stack optimization framework as shows in Fig.~\ref{fig:full-stack}. First, they use the quantization-aware training for power-of-two weights. Then, use the column combining technique~\cite{kung2019packing} to make the weights of kernel be structured and sparsity. Then, design a systolic array~\cite{kung1982systolic} based accelerator with selector accumulator to inference the model. The powerful method in \cite{mcdanel2019full} is use the power-of-two weights to perform inference. In the hardware view, when you times power-of-two number, it only need to increase or decrease zero from the original number. Because of this method, the accelerator do not need the MAC unit (multiply accumulate), just use the selector accumulator to perform inference. However, it can only inference the customized model with power-of-two weights.

\begin{figure}[h]
    \centering
    \includegraphics[width=8cm]{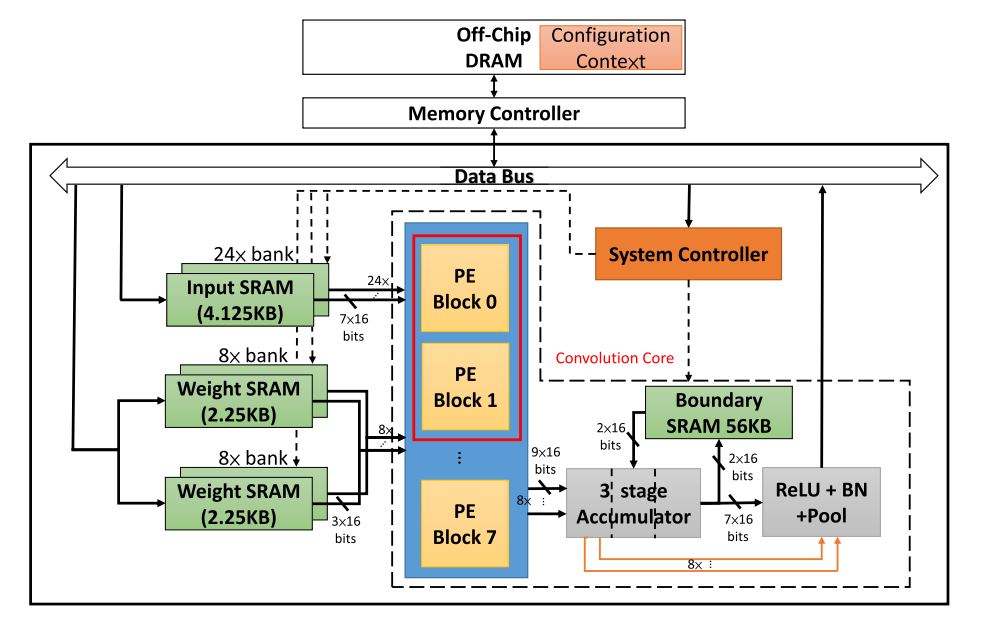}
    \caption{Overview of the vector-wise CNN accelerator.~\cite{chang2019vwa}}
    \label{fig:VWA}
\end{figure}

In \cite{chang2019vwa}, they proposed a vector-wise CNN accelerator as shows in Fig.~\ref{fig:VWA}. They use vector-wise input approach to maintain the high hardware utilization. And use kernel decomposition~\cite{lin2017data} and interleaved input for handling different convolution types to maintain the high hardware utilization. Then, uses three-stage accumulator to build the reconfigurable accumulators. But it did not have enough computation performance to inference the major models.

\subsection{Embedded CNN Structure}
\label{s:embedded_CNN}

\begin{figure}[h]
    \centering
    \includegraphics[height=7cm]{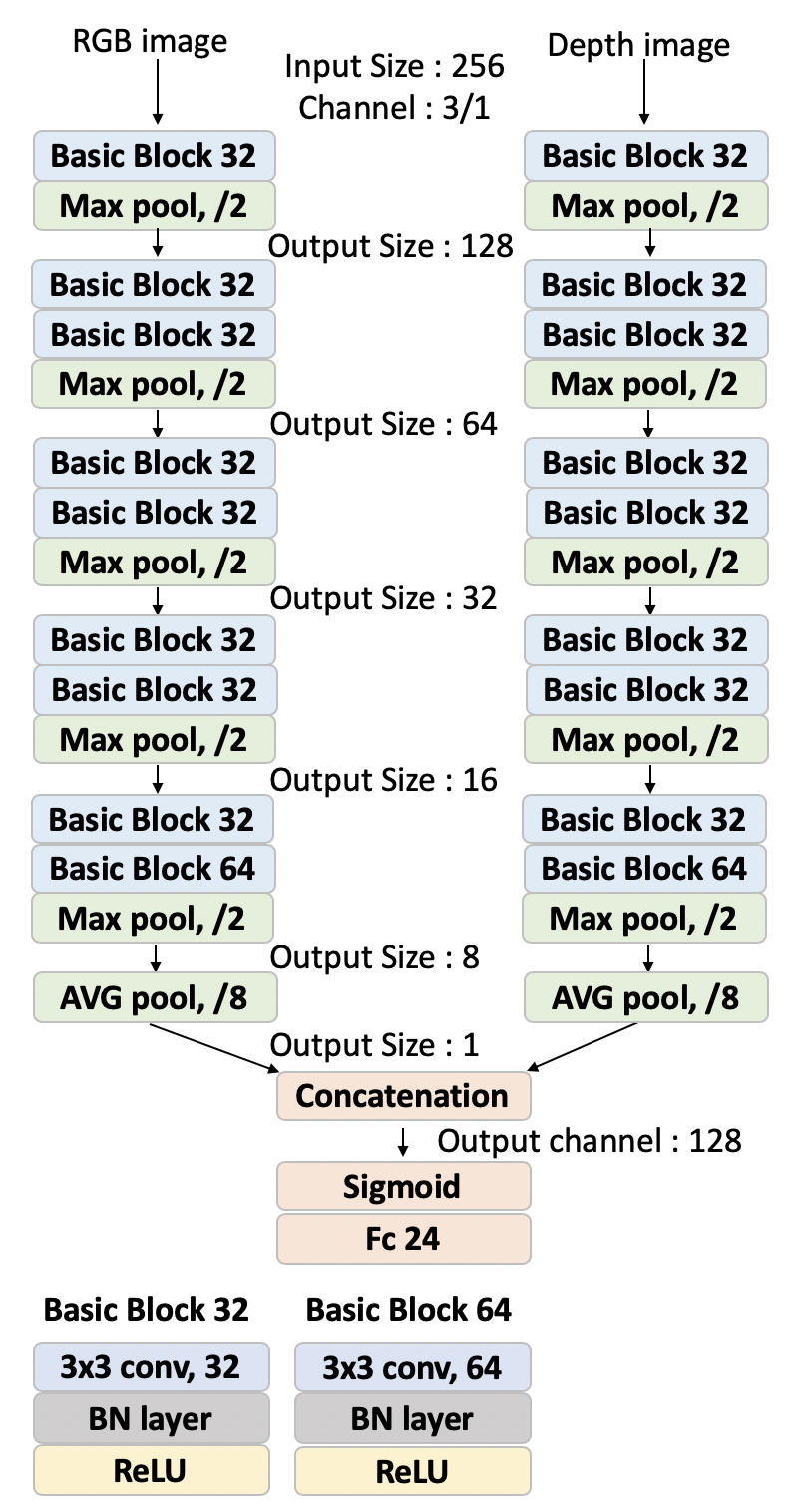}
    \caption{Overview of the model RGBD eCNN.~\cite{wang2020fast}}
    \label{fig:model_ECNN}
\end{figure}

In \cite{wang2020fast}, they proposed an embedded CNN structure , RGBD eCNN, as shows in Fig.~\ref{fig:model_ECNN}. This model with only nine convolution layers to reduce the computation complexity of inference. And input feature map size of each layers are power-of-two is more easier to perform the input feature map segmentation. Also the input and output channel numbers are power-of-two, which are more suitable to do the tiling technique. Then, use max pooling in the earlier layers to reduce the feature map size and reduce the times of tiling. Because of the above reasons, we choose the RGBD eCNN as the target model to design of the proposed accelerator.

\section{Proposed Methods}
\label{sec:proposed_methods}

In this chapter, we describe the proposed accelerator system architecture and all methods we use for reduce DRAM data access amount and maintain the high utilization of hardware resource. In section~\ref{s:proposed_methods}, we present our proposed methods and techniques to improve the energy efficiency, keep the high hardware utilization and make our system more flexible to support different kernel sizes and strides. In section~\ref{s:system_overview}, we give an overview of the proposed system architecture and talk about our design concept. In section~\ref{s:hardware_unit_design}, we introduce the important hardware component of our system.

\subsection{Proposed Methods}
\label{s:proposed_methods}

\begin{figure}[h]
    \centering
    \includegraphics[width=8cm]{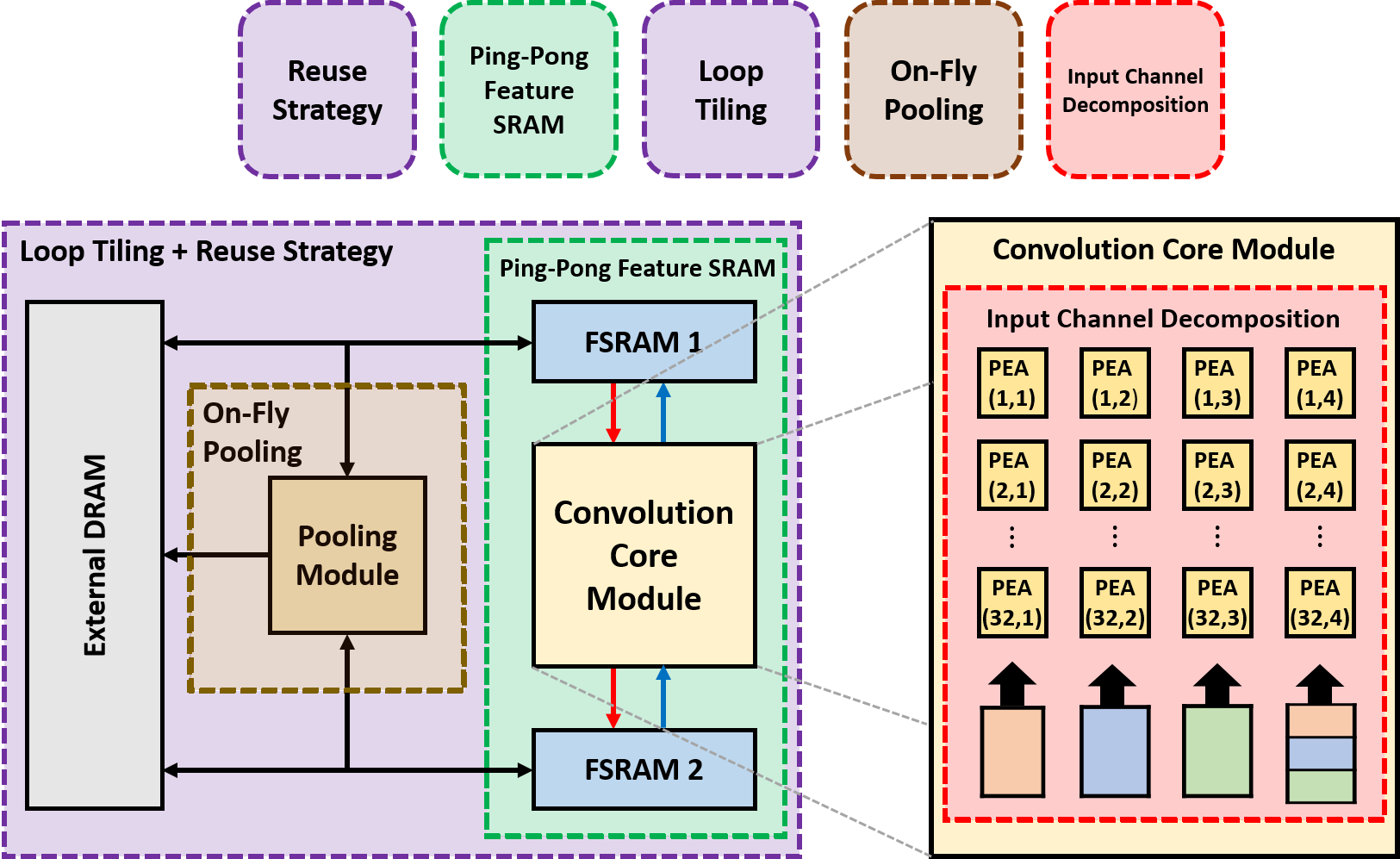}
    \caption{System overview of the proposed accelerator and methods.}
    \label{fig:proposed_methods}
\end{figure}

As shown in Fig.~\ref{fig:proposed_methods}, proposed accelerator has five major methods, reuse strategy, ping-pong feature SRAM, loop tiling, on-fly pooling and input channel decomposition.

We notice that the key factor to improve the computation performance is to improve the hardware utilization of the convolution core module as much as possible. Therefore, we proposed ring streaming dataflow and input channel decomposition techniques to maintain the high utilization. On the other hand, we proposed the loop tiling and reuse strategy to optimize our parameters of hardware resource and reduce the data access amount.

\subsubsection{Loop Tiling}
\label{s:loop_tiling}
The tiling technique is to parallelize the computation, it aim at high utilizing the hardware resources and has been proposed in \cite{hsu2020essa} by us before, but has improvement in this thesis. We are not able to compute full convolution in the same time, so we need to split each the channels of input feature map and kernel numbers. Therefore, we propose three tiling factor $T_m$, $T_n$ and $T_{size}$ to split the whole convolution to several parts. $T_m$ is the number of computed output channel each time, $T_n$ is the number of computed input channel each time and $T_{size}$ is the size of computed size of input channel each time.

Different tiling factors pairs will produce different computation performance and data access amount. The appropriate tiling factors pairs can increase the computation efficiency and data reuse, so we integrate the reuse strategy in section~\ref{s:reuse_strategy} and simulate all possible factor pairs to find out the highest computation performance with the lower data access amount with a program.

\subsubsection{Reuse Strategy}
\label{s:reuse_strategy}

The purpose of the reuse strategy is to thrift the data access amount with external DRAM. We design two kinds of reuse strategies, which are input reuse and output reuse. We formulate the data access amount as:

\begin{equation}
\label{eq:data_access_amount}
\resizebox{0.5\textwidth}{!}{$
Data Access Amount = {Data_{in}} \times{Time_{in}} \times {Data_{weight}} \times{Time_{weight}} \times {Data_{out}} \times{Time_{out}}$}
\end{equation}

In eq.\ref{eq:data_access_amount}, ${Data_{in}}$, ${Data_{weight}}$ and ${Data_{out}}$ represent the external data access amount in each tiling time of input feature maps, kernel weights and output feature maps. ${Time_{in}}$, ${Time_{weight}}$ and ${Time_{out}}$ represent the trip count to tiling the whole convolution of input feature maps, kernel weights and output feature maps.

\begin{figure*}[h]
    \centering
    \includegraphics[width=16cm]{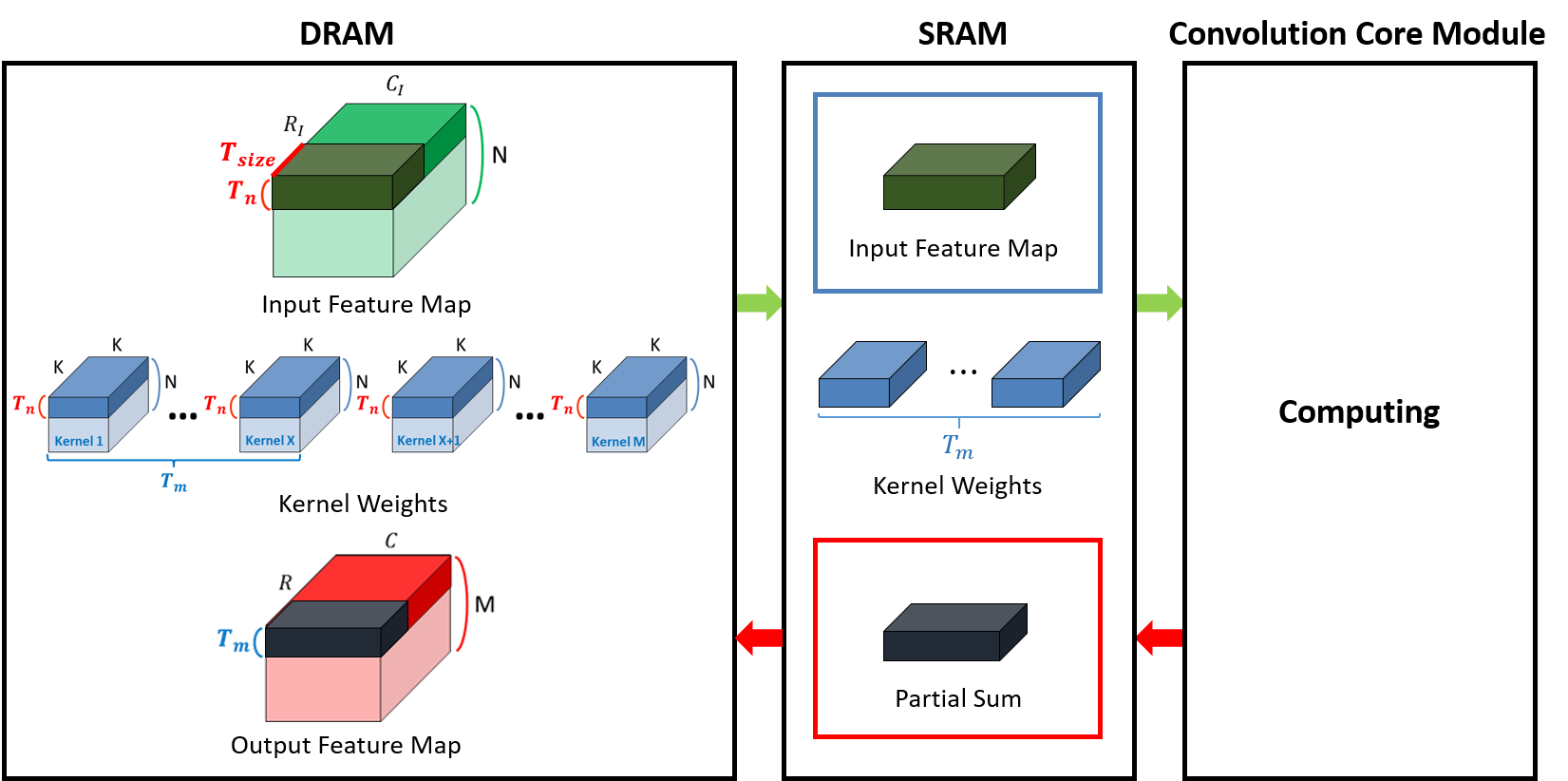}
    \caption{Different reuse parts in input (blue box) and output (red box) feature map.}
    \label{fig:reuse_strategy}
\end{figure*}

As shown in Fig.~\ref{fig:reuse_strategy}, Reuse strategy can be roughly divided into two types, input reuse and output reuse. Input Reuse stored input feature map (blue box) in SRAM and stored the partial sum back to DRAM and output Reuse stored the partial sum (red box) in SRAM to reuse.

For the input reuse strategy, we can calculate the data access amount with eq.\ref{eq:data_access_amount} and the factors in this equation can be formulate as: 
\begin{equation}
\left\{
             \begin{array}{lr}
             {Data_{in}} = T_n \times {(S \times {Feature_{size}+K-S})}^{2} \\
             {Time_{in}} = \frac{N}{T_n} \\
             {Data_{weight}} = K \times K \times T_m \times T_n \\
             {Time_{weight}} = \frac{M}{T_m} \times \frac{N}{T_n} \\
             {Data_{out}} = {Feature_{size}}^{2} \times T_m (non-pooling) \\
             {Data_{out}} = {Feature_{size}}^{2} \times T_m \div 4 (pooling)\\
             {Time_{out}} = 2 \times \frac{N}{T_n} \times \frac{M}{T_m},
             \end{array}
\right.
\label{eq:Input_Reuse}
\end{equation}

For the output reuse strategy, we can calculate the data access amount with eq.\ref{eq:data_access_amount},too. However, the factors in this equation will be formulate as: 
\begin{equation}
\left\{
             \begin{array}{lr}
             {Data_{in}} = T_n \times {(S \times {Feature_{size}+K-S})}^{2} \\
             {Time_{in}} = \frac{M}{T_m} \times \frac{N}{T_n} \\
             {Data_{weight}} = K \times K \times T_m \times T_n \\
             {Time_{weight}} = \frac{M}{T_m} \times \frac{N}{T_n} \\
             {Data_{out}} = {Feature_{size}}^{2} \times T_m (non-pooling) \\
             {Data_{out}} = {Feature_{size}}^{2} \times T_m \div 4 (pooling) \\
             {Time_{out}} = \frac{M}{T_m}
             \end{array}
\right.
\label{eq:Output_Reuse}
\end{equation}

The different between these two kinds of reuse strategy is ${Time_{in}}$ and ${Time_{out}}$. Because of the input reuse strategy is to reuse the input feature map in convolution and store the partial sum back to the external DRAM and the output reuse strategy is to Store the partial sum of convolution in on-chip SRAM. Therefore, ${Time_{out}}$ in eq.\ref{eq:Input_Reuse} of input reuse strategy have a factor 2 is for both read and write the partial sum between external DRAM and on-chip SRAM of output feature map.

\subsubsection{Tiling Factor}
\label{s:tiling_factor}

We use the loop tiling technique and output reuse strategy to research optimal tiling factors with the limitation of MAC number smaller than 1152. Then use the program to find the pair of factors, have highest computation performance with the lowest data access amount of each convolution layers in RGBD eCNN. As shown in table~\ref{tb:tiling_factors}.

\begin{table}[h]
\centering
\caption{Optimal tiling factors of each layers in RGBD eCNN.}
\label{tb:tiling_factors}
\begin{tabular}{|c|c|c|c|c|c|}
\hline
          & Tm & Tn & Cycle  & Performance   (GOPs) & Data   Access (MB) \\ \hline
Layer   1 & 32 & 3  & 65538  & 863.97               & 2.191265           \\ \hline
Layer   2 & 32 & 4  & 131088 & 1152                 & 1.024536           \\ \hline
Layer   3 & 32 & 4  & 131088 & 1152                 & 1.024536           \\ \hline
Layer   4 & 32 & 4  & 32784  & 1152                 & 0.266724           \\ \hline
Layer   5 & 32 & 4  & 32784  & 1152                 & 0.266724           \\ \hline
Layer   6 & 32 & 4  & 8208   & 1152                 & 0.075317           \\ \hline
Layer   7 & 32 & 4  & 8208   & 1152                 & 0.075317           \\ \hline
Layer   8 & 32 & 4  & 2064   & 1152                 & 0.026489           \\ \hline
Layer   9 & 32 & 4  & 4128   & 1152                 & 0.052979           \\ \hline
Total     &    &    &        &                      & 5.003887           \\ \hline
\end{tabular}
\end{table}

\begin{figure}[h]
    \centering
    \includegraphics[width=8cm]{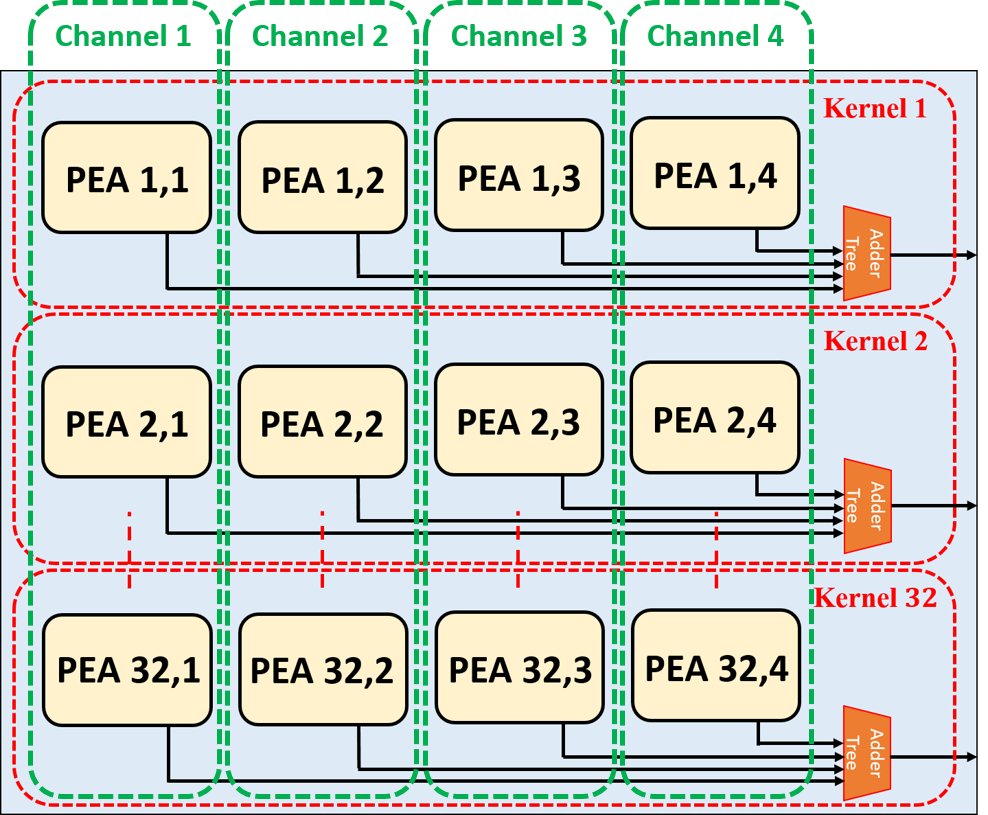}
    \caption{Mapping convolution of each tiling to CCM.}
    \label{fig:CCM_tiling}
\end{figure}

The proposed convolution core module (CCM) contains 32 rows and 4 column as Fig.~\ref{fig:CCM_tiling}. Each row of CCM calculates the partial sum with 4 filters in the same kernel and each column of CCM calculates the 1 input channel with different kernel. Sum up the results from the each row of process element array (PEA) as partial sum of one output channel.

For the bank number of each SRAM, according to the tiling factors, each SRAM in data unit needs 32 banks to store the data by channel parallelly. In addition, the proposed data init use two same size feature SRAM (FSRAM) to build up ping-pong FSRAM.

\subsubsection{Ring Streaming Dataflow}
\label{s:ring_streaming_dataflow}

\begin{figure}[h]
\begin{center}
\subfigure[Traditional streaming dataflow] {
	\includegraphics[width=4cm]{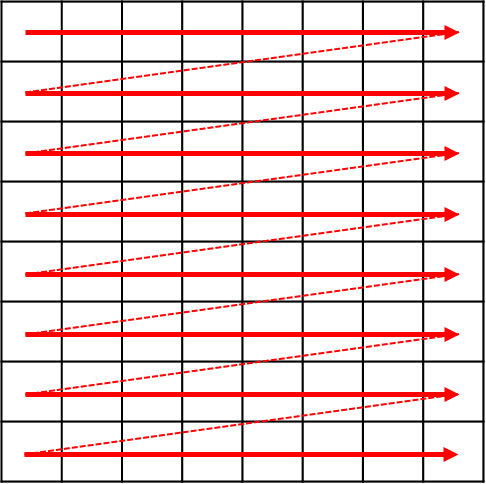}
	\label{fig:traditional_dataflow}
}
\subfigure[Proposed ring steaming dataflow] {
	\includegraphics[width=4cm]{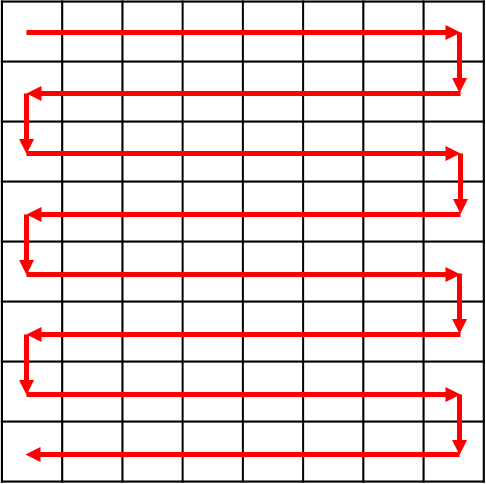}
}
\caption{The dataflow of (a) traditional streaming and (b) proposed ring streaming on output feature map}
\label{fig:two_kinds_dataflow}
\end{center}
\end{figure}

In order to reduce the data loading time and increase hardware utilization and computation performance, we propose the ring streaming dataflow. As shown in Fig.~\ref{fig:two_kinds_dataflow}, transitional streaming dataflow process each row of output feature map from left to right sequentially, but our proposed ring streaming dataflow process the output feature map by ring. That means the ring streaming dataflow process the output feature map from the leftmost to the rightmost then shift one pixel below then from the rightmost to the leftmost repeatedly. 

\subsubsection{Input Channel Decomposition}
\label{s:input_channel_decomposition}

Convolution layer 1 with only three input channels is a big issue for CNN accelerator utilization. In order to maintain high utilization in layer 1, we propose the input channel decomposition technique. Input channel decomposition can be divided into two steps, using the size tiling factor that we introduce in section~\ref{s:loop_tiling} to split convolution into several tasks then assign these tasks to each process element array column as evenly as possible.

\begin{figure}[h]
    \centering
    \includegraphics[width=8cm]{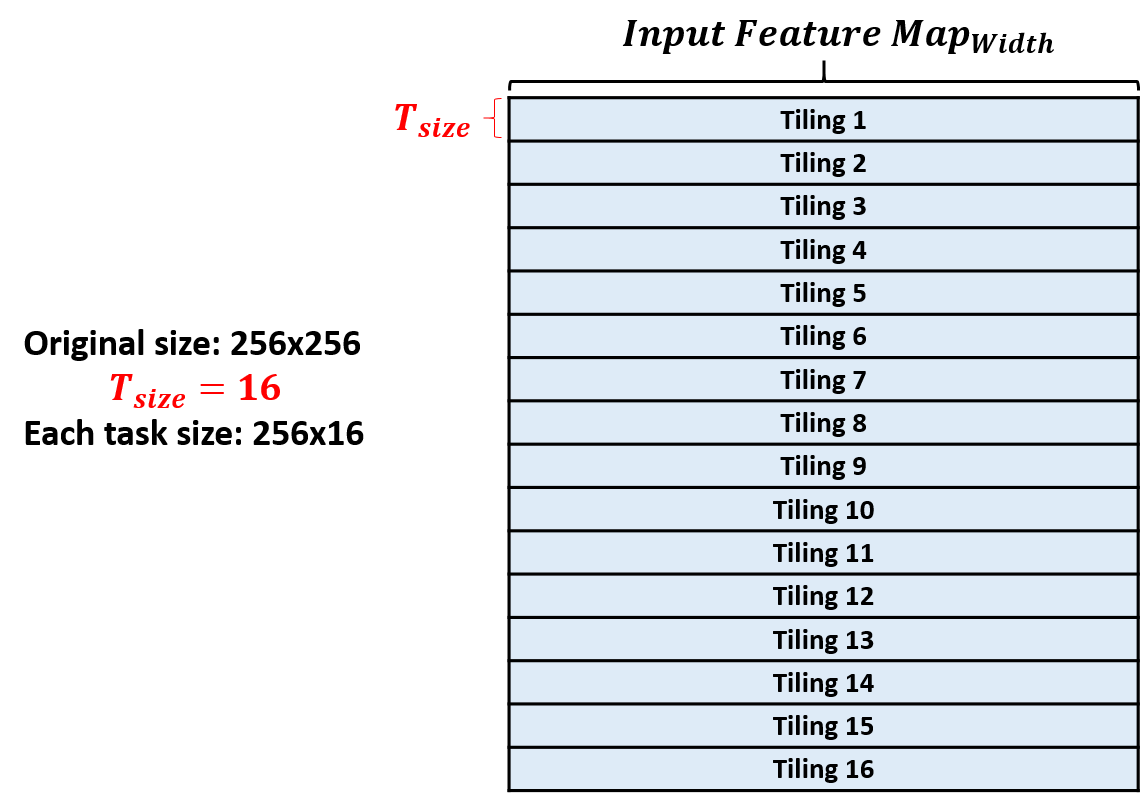}
    \caption{Input feature map from RGBD eCNN~\cite{wang2020fast} layer 1 split into 16 tasks.}
    \label{fig:input_channel_decompostion_one_map}
\end{figure}

Here, we take model RGBD eCNN~\cite{wang2020fast} for example, the input feature map resolution of RGBD eCNN layer 1 is 256x256 and the $T_{size}$ will be 16 to fit the SRAM size. Thus, the input feature map of each channel can be split into 16 tiling tasks as Fig.~\ref{fig:input_channel_decompostion_one_map} and RGBD eCNN layer 1 input feature map have three channels, so it will be 48 tiling tasks at all.

\begin{figure}[h]
    \centering
    \includegraphics[width=8cm]{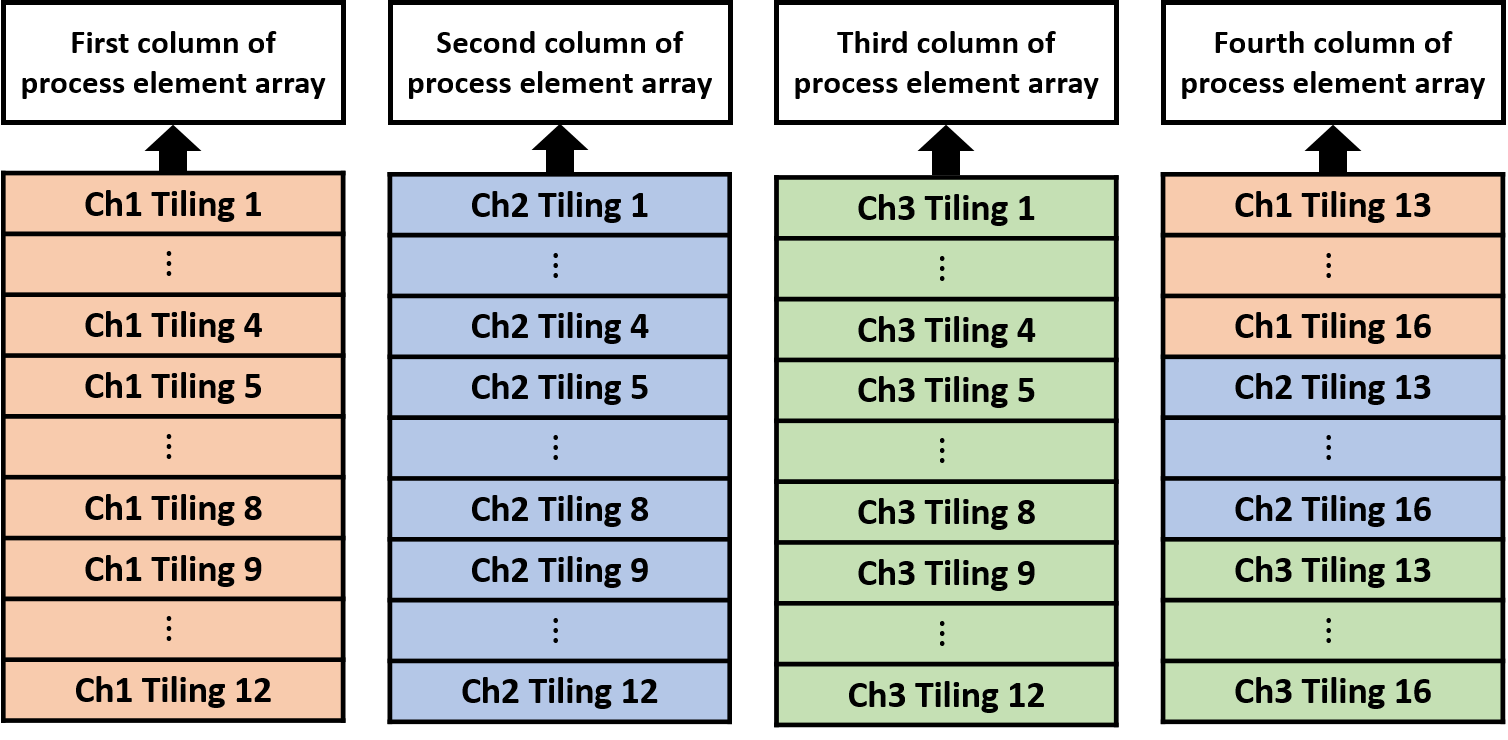}
    \caption{Proposed input channel decomposition in RGBD eCNN layer 1.}
    \label{fig:input_channel_decompostion}
\end{figure}

As shown in Fig.~\ref{fig:input_channel_decompostion}, there are four columns process element array (PEA) in our proposed convolution core module. Each column of PEA can process 1 input channel with 32 output channels. If we don't use the input channel decomposition technique, the fourth column of PEA will idle in whole layer 1 convolution time. With the proposed input channel decomposition technique tiling tasks 13 to 16 from channel 1 to channel 3 can be assign to the fourth column of PEA and process the convolution in the same time. 

\subsection{System Overview}
\label{s:system_overview}

As shown in table~\ref{tb:qutize}, RGBD eCNN has been quantized from 32 bits to 8 bits to reduce the data access amount and the SRAM size. And we quantized the VGG-16 from 32 bits to 8 bits and only decreased 0.54\% accuracy. Therefore, we set the bit width of convolution of the proposed accelerator to 8 bits.

\begin{table}[h]
\centering
\caption{Data access amount comparison on model RGBD eCNN.}
\label{tb:qutize}
\begin{tabular}{|c|c|c|c|}
\hline
CNN type  & Precision & Accuracy & Accuracy   lose \\ \hline
RGBD eCNN & 32 Bits   & 99.89\%  &                 \\ \hline
RGBD eCNN & 8   Bits  & 99.79\%  & -0.1\%          \\ \hline
VGG-16    & 32 Bits   & 70.5\%   &                 \\ \hline
VGG-16    & 8   Bits  & 69.96\%  & -0.54\%         \\ \hline
\end{tabular}
\end{table}

\begin{figure}[h]
    \centering
    \includegraphics[width=8cm]{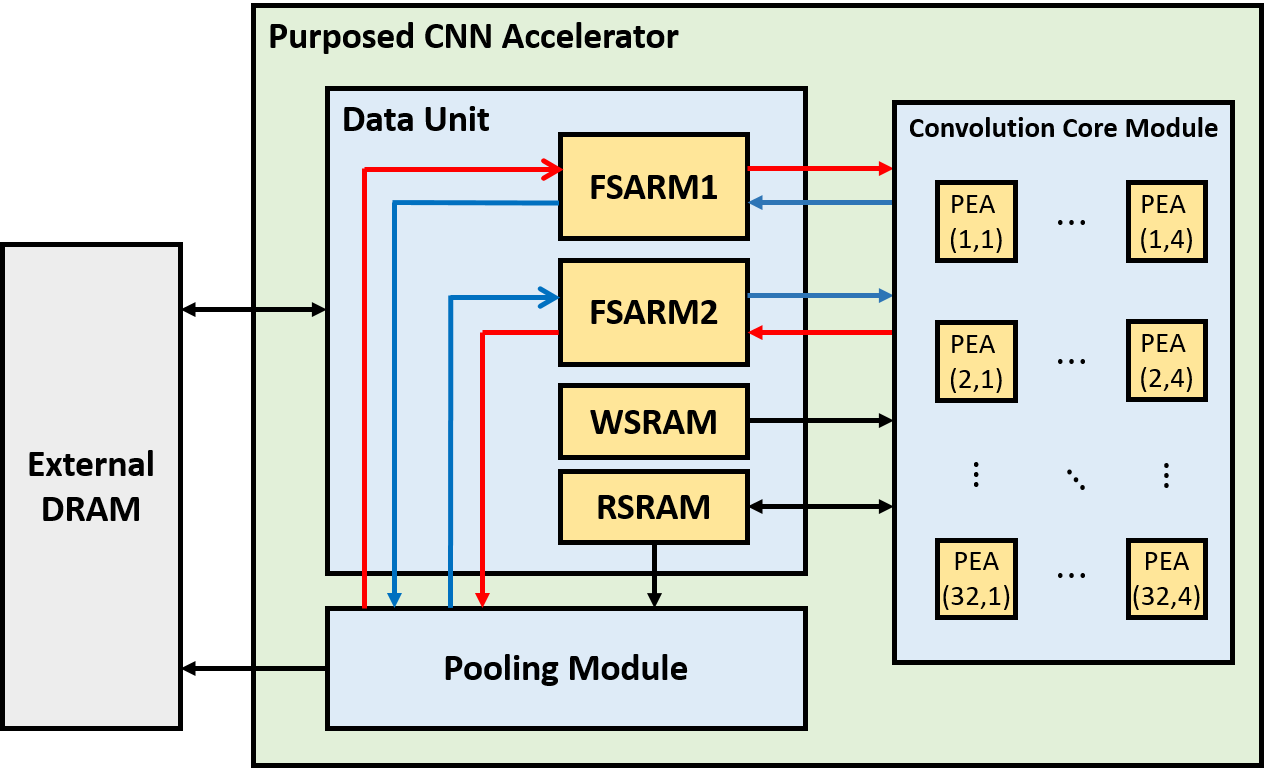}
    \caption{Proposed system overview.}
    \label{fig:system_overview}
\end{figure}

Fig.~\ref{fig:system_overview} gives an overview of the proposed accelerator. The proposed accelerator consist of three major components, data unit, convolution core module and pooling module. Input feature map and kernel weights transferred from external DRAM to data unit, which consist of SRAM controller for handle dataflow and data placement, and multi-bank feature, weight and reuse SRAM for saving data. We have two same size feature SRAM to build up a ping-pong buffer, each feature SRAM has 32 banks to save different channels of input feature maps or output feature maps. Weight SRAM also is built with 32 banks in order to store 32 sets of kernels. Reuse SRAM is used to store the reused input feature maps and the output feature maps, which is waiting for pooling. There is no need to pre-load whole input feature map from the DRAM, but the top two rows. Meanwhile, kernel weights have been loaded to weight SRAM. After all of the data that convolution needed are loaded in data unit, we can start to transfer kernel weights to convolution core module. When kernel weights have been written, then we begin conveying input feature map and computing the output feature map. We can calculate the sum of 32 output feature maps in each cycle and we write back the value into another feature SRAM, which is empty. 

Take an example for the flow of convolution with red line in Fig.~\ref{s:system_overview}. Kernel and input feature map will be loaded to WSRAM and FSRAM 1 and transfer kernel and data to CCM to perform convolution. Each cycle CCM write a partial sum back to FSRAM 2 then after the convolution if there is a pooling layer after current layer, FSRAM 2 will transfer output feature map to pooling module to perform the max pooling. Finnaly, pooling module will write the results back to FSRAM1 or DRAM. 

There are three issues for our CNN accelerator design. First, the computation performance of the proposed accelerator system should be able to support the state-of-the-art models like VGG-16~\cite{simonyan2014very}, MobileNet v1~\cite{howard2017mobilenets} inference in real-time. The computation performance can be formulized as:

\begin{equation}
\label{eq:computation_performance}
\resizebox{0.5\textwidth}{!}{$
Computation Performance = 2 \times \# of MAC \times {\color{red}Hardware  Utilization}\times Frequency$}
\end{equation}

In equation \ref{eq:computation_performance}, we need a factor, 2, to multiply numbers of MAC is that because when executing the convolution computation, there are one multiply with one accumulate, the total numbers of computation is 2. In the proposed accelerator design, the numbers of MAC and the clock frequency are unchangeable. As a result, we can notice the only flexible variable is the hardware utilization.

Second, we want our proposed accelerator can be placed in edge devices in order to process the image classification tasks. Therefore, the higher energy efficiency will be the primary consideration. Energy efficiency can be formulized as:
\begin{equation}
\label{eq:energy_efficiency}
Energy Efficiency = \frac{Computation Performance}{Energy Consumption}
\end{equation}
We discover the most important thing for optimizing the energy efficiency is to reduce the power consumption of our proposed accelerator. Hence, the excessive energy consumption part in convolution computation is the external DRAM access. That is to say, if we can reduce the external DRAM access amount when computing convolution, we can obtain higher energy efficiency. And the greater hardware utilization can also reduce the inference latency, this would further be beneficial to reduce energy consumption.

Third, with the two issues mentioned above, we can find out that hardware utilization is the most predominant issue. The reason that the current accelerators can not maintain high utilization is there are too many different convolution types of the state-of-the-art CNN models. Therefore, our proposed accelerator design needs to meet the requirement of flexible convolution. The following three convolution types are supported by the proposed accelerator, zero padding in input feature map, stride two in convolution layer and two prevalent kernel size.

For the design of zero padding in input feature map, we design an innovative SRAM placement, Double Pixels per Row (DDPR), which can handle padding more suitable to reduce loading latency and memory space to store zero. Next, the solution for striding two in convolution layer is that we use dual-port SRAM and Double Pixels per Row to conquer this question with no negative affection. Lastly, the issue of different kernel sizes, the proposed accelerator design uses 1x1 computing component for basic block. So we can process the convolution of 1x1 kernel size without sacrificing a great quantity of hardware utilization.

\subsection{Hardware Unit Design}
\label{s:hardware_unit_design}

In this section, the proposed CNN accelerator hardware specific will be introduced. The proposed hardware consists of three main modules, data unit, convolution core module and pooling module. We give the design idea and the detail information of each module in the subsections below.

\subsubsection{Convolution Core Module}
\label{s:convolution_core_module}

\begin{figure}[h]
    \centering
    \includegraphics[width=8cm]{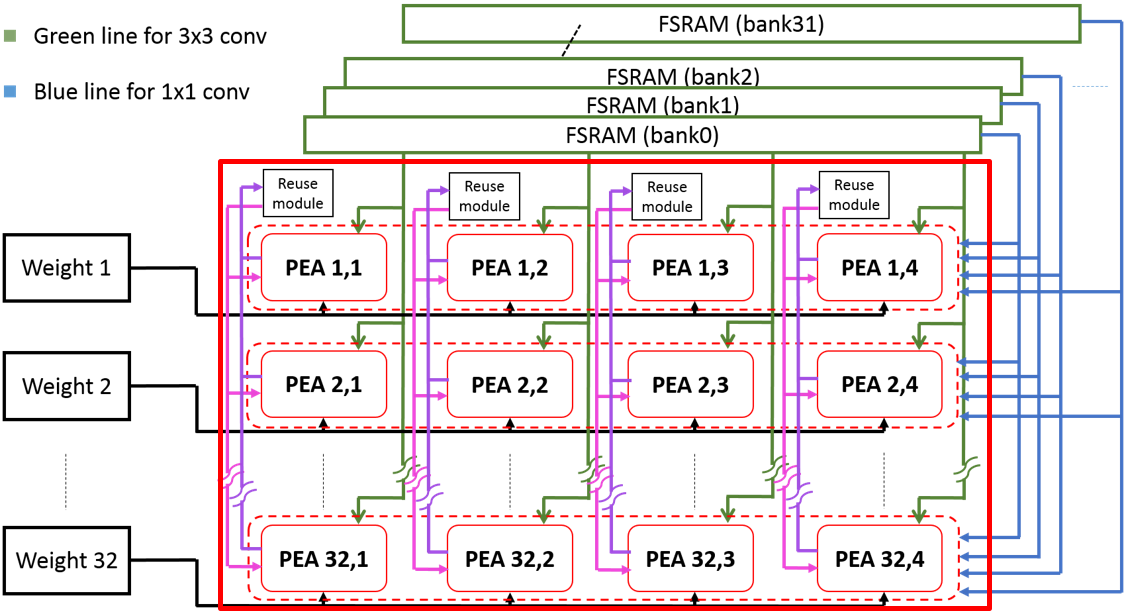}
    \caption{Overview of the proposed convolution core module.}
    \label{fig:CCM}
\end{figure}

Inside the red box of the Fig.~\ref{fig:CCM} shows the overview of the proposed convolution core module. The proposed convolution core module (CCM) consists of two parts, 128 process element arrays (PEA) and four reuse modules. There are two different lines connect from the feature SRAM to process element arrays, green line and blue line. Green line for kernel size 3x3 convolution and blue line for kernel size 1x1 convolution. And the reuse module can store the data, which can be reuse in convolution.

\begin{figure}[h]
    \centering
    \includegraphics[width=8cm]{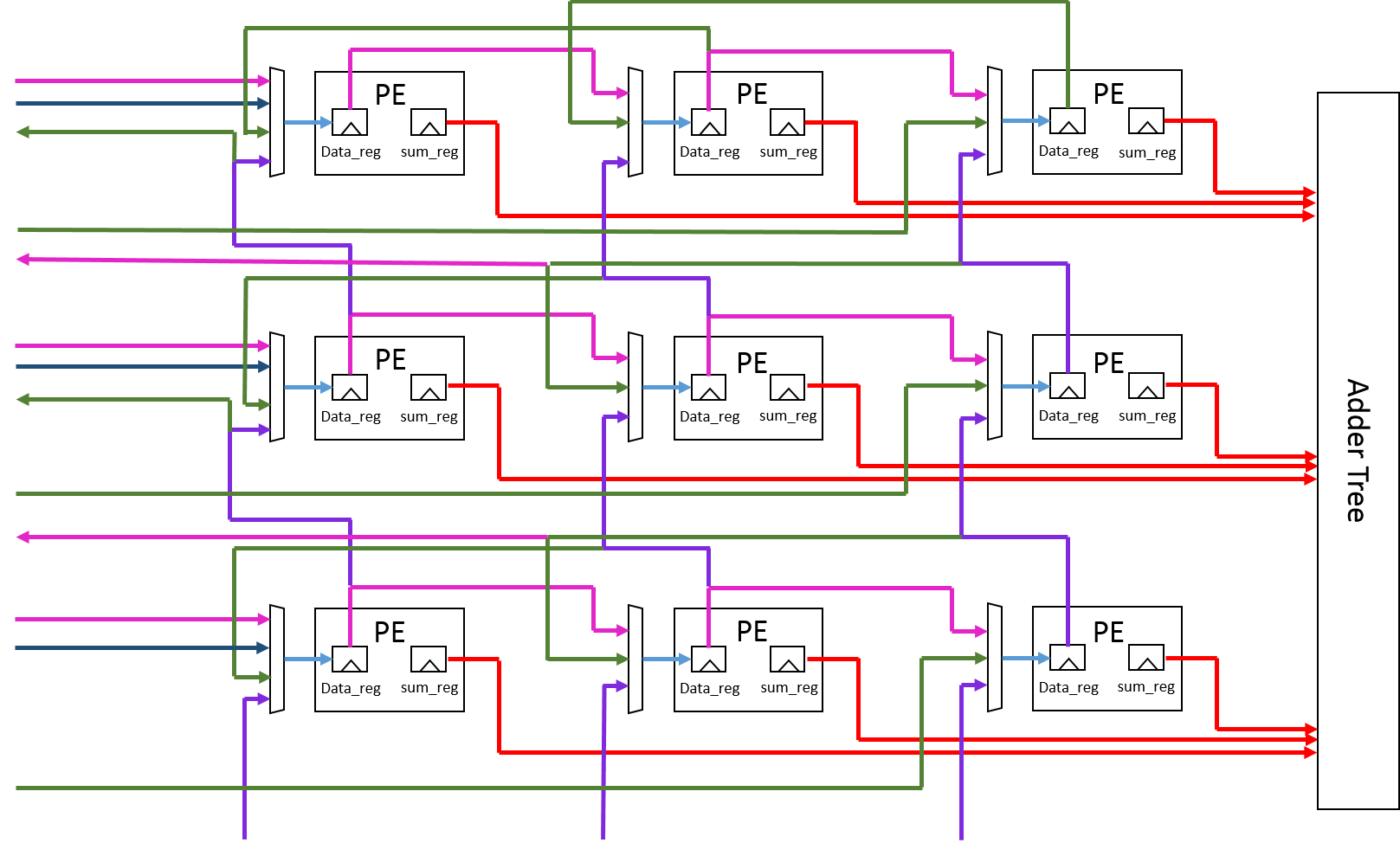}
    \caption{The architecture of the proposed process element array.}
    \label{fig:PEA}
\end{figure}

Fig.~\ref{fig:PEA} shows the architecture of the proposed process element array (PEA), it consist of nine process element (PE). Each process element can compute convolution with kernel size 1x1, and using a adder tree to sum up results from nine process element.

\begin{figure}[h]
    \centering
    \includegraphics[width=9cm]{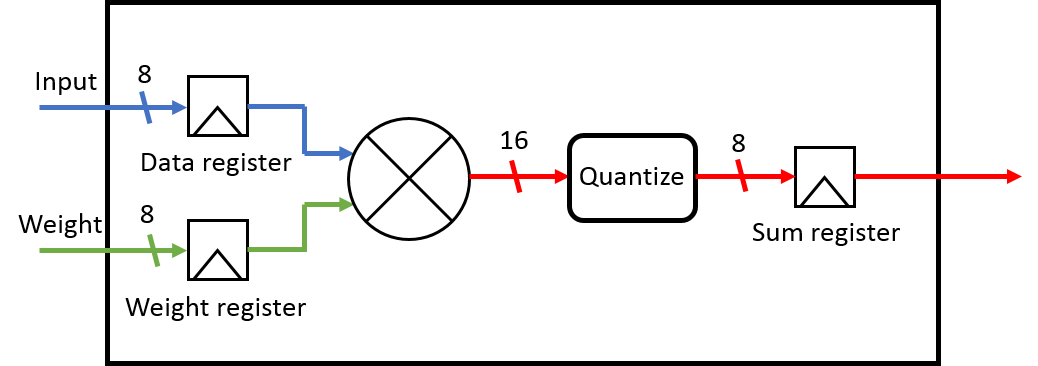}
    \caption{The architecture of the proposed process element.}
    \label{fig:PE}
\end{figure}

Fig.~\ref{fig:PE} shows the architecture of the proposed process element (PE), each PE can execute a 1x1 convolution with Multiply the input data and weight ,then quantize the result to 8 bits and write in the sum register.

In order to realize the proposed ring streaming dataflow, we need to design four different kinds of dataflow. Our proposed process element array can shift in three direction, left, right and up. Each direction has the interrelated dataflow in the design of our process element array. 

As shown in Fig.\ref{fig:PEA}, for the front shifting dataflow, only use for convolution first three rows of input feature map. When executing the front shifting, the data stored in the data register of process element will transfer to the process element at right hand side. And the data, which store in the right most process element will be transfer to the reuse module for the inverted direction shifting. Since there have no data in the reuse module, so we need three pixels from the feature SRAM in each cycles to calculate one output pixel. For the left and right shifting, When executing the shifting, the data stored in the data register of process element will transfer to the process element at left and right hand side, respectively. And the input data of first two rows of process element array will be provided by the reuse module. For the up shifting, the data stored in the data register of process element will transfer to the above process element.

\begin{figure}[h]
    \centering
    \includegraphics[width=8cm]{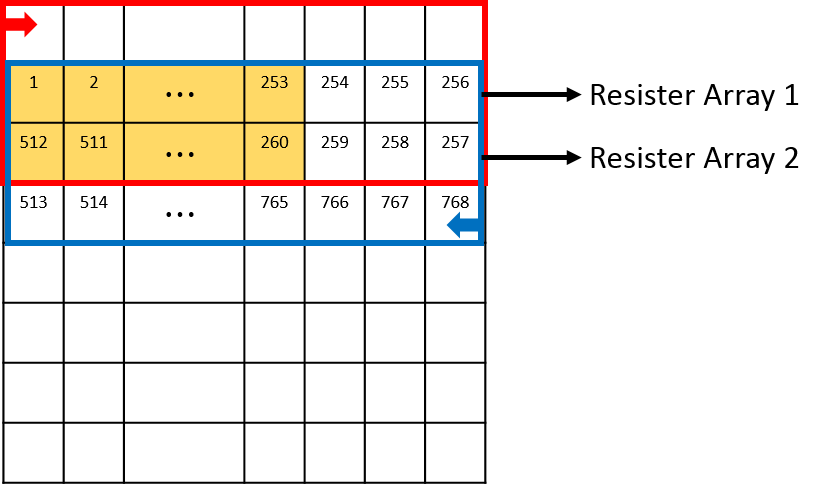}
    \caption{Yellow block are the pixels of input feature map that can be reused.}
    \label{fig:reuse_part}
\end{figure}

As shown in Fig.~\ref{fig:reuse_part}, the pixels inside the red box are use for the front shifting, and the pixels inside the blue box are use for the left shifting. The overlap of this two shifting (yellow part) are the pixels, which can be reused in current left shifting (blue box). Therefore, we store them in the reuse module, and it can reduce 66\% data access amount of the on-chip SRAM.

The proposed reuse module consists of two register arrays, and each array can store 222 pixels. The reason of the pixel number can be split as (221+1) pixels, 221 pixels for pixels can be reuse in each row of feature map, and 1 for stagger the the read and write operation. The data, which can be reused, will be transferred from process element array. And the reuse data will be read out in the next inverted shifting.

\begin{figure}[h]
    \centering
    \includegraphics[width=8cm]{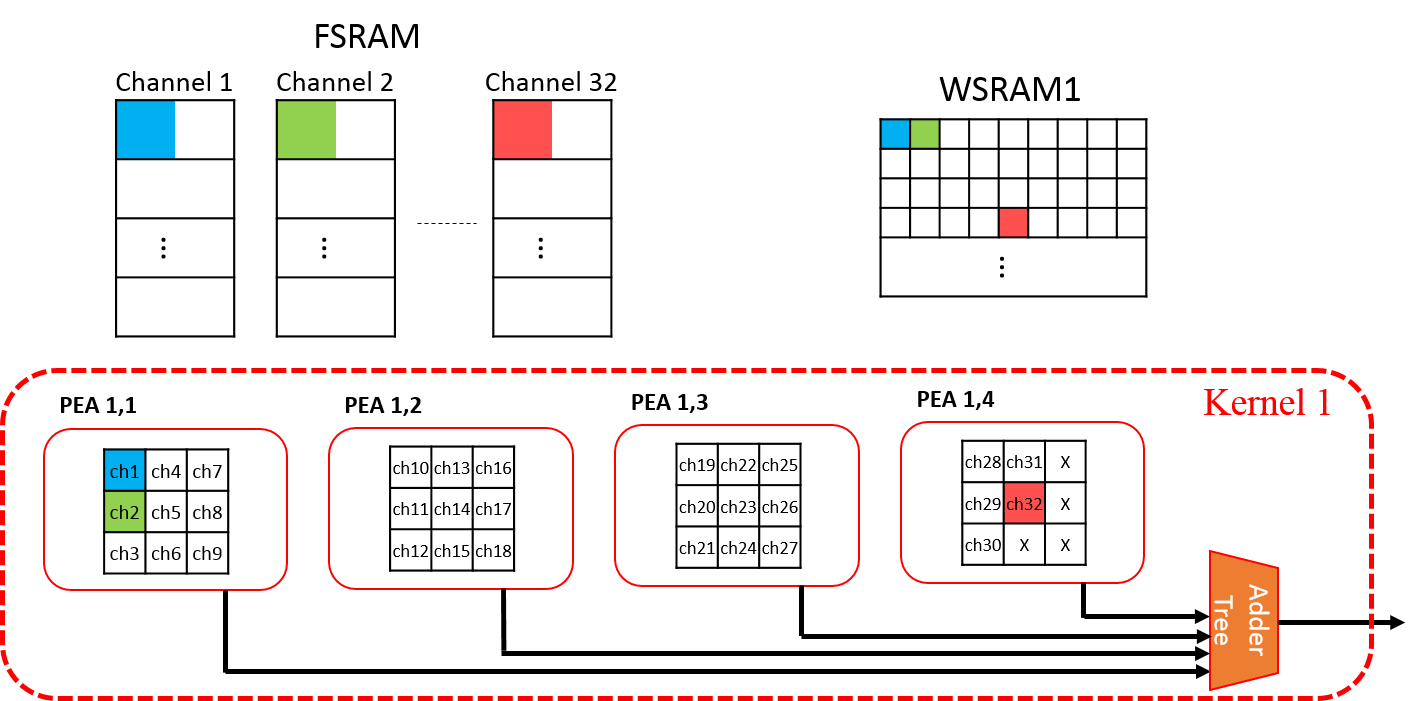}
    \caption{1x1 convolution dataflow with 1 output channel.}
    \label{fig:1x1_conv_1ch}
\end{figure}

Our proposed convolution core module uses 1x1 computing process element for basic block. Therefore, we can change to a suitable dataflow to reach a higher hardware utilization in the 1x1 convolution. As Fig.~\ref{fig:1x1_conv_1ch} shows, the convolution data flow with 1 output channel. First, fetch the same coordinate pixel in 32 input channels to PEs, then fetch 32 1x1 filters from same kernel to PEs to perform convolution. Finally, The sum of 32 results from PEs is equal to 32ch to 1ch 1x1 convolution.

\subsubsection{Data Unit}
\label{s:data_unit}

\begin{figure}[h]
    \centering
    \includegraphics[height=7cm]{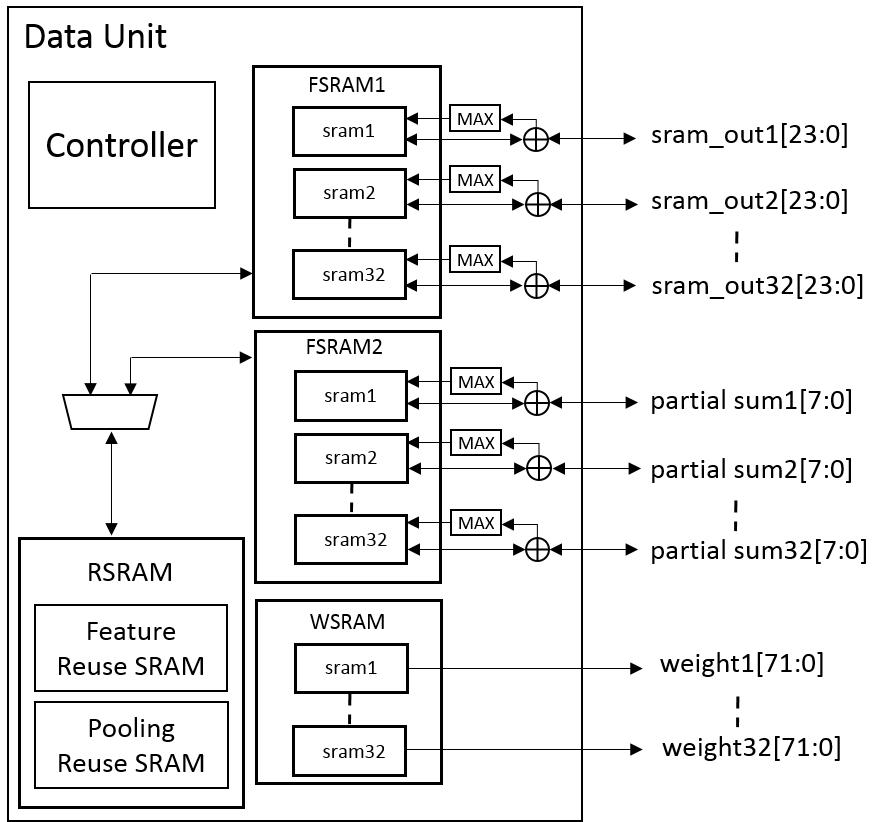}
    \caption{The overview of proposed data unit.}
    \label{fig:Data_unit}
\end{figure}

As shown in Fig.~\ref{fig:Data_unit}, our proposed data unit consist of one controller, two feature SRAMs (FSRAM) to build up ping-pong feature SRAM (PPFS), one weight SRAM (WSRAM) and one reuse SRAM (RSRAM). 

Each feature SRAM consists of 32 dual-ports SRAM banks to store the input feature map parallelly by channel. Weight SRAM consists of 32 SRAM banks to store the kernel weights parallelly by channel, too. Reuse SRAM consists of two parts of SRAM banks, feature reuse SRAM and pooling reuse SRAM, which has different purposes. At last, we use the controller to decide the dataflow in different situations.

\begin{figure}[h]
    \centering
    \includegraphics[width=8cm]{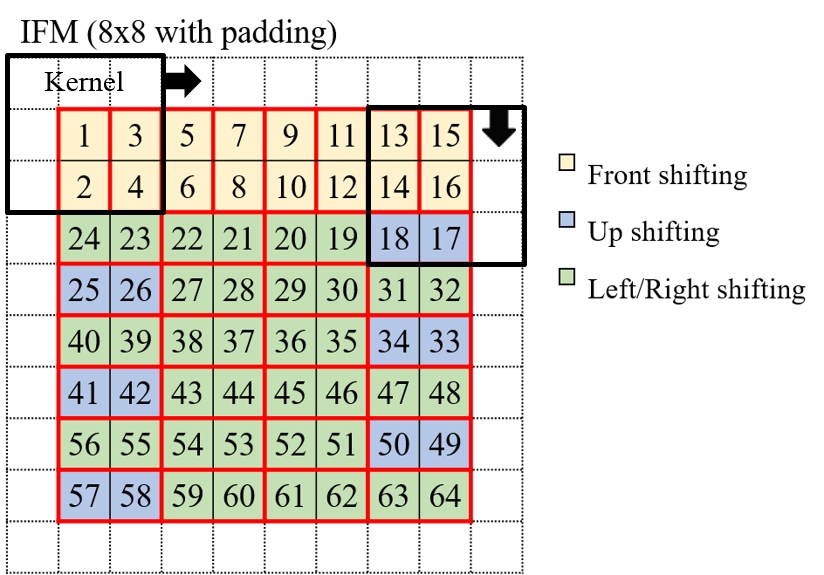}
    \caption{Data placement of the proposed feature SRAM.}
    \label{fig:FSRAM_data_placement}
\end{figure}

In order to increase the computation performance of the proposed accelerator, we propose the double pixels per row (DPPR). As shown in Fig.~\ref{fig:FSRAM_data_placement}, two pixels of input feature map, which framed with red squares, we store them in the same row of feature SRAM. In addition, the purposed accelerator supports the zero padding, but do not need to store padding 0s in the feature map SRAM is because the controller in data unit will fetch up the padding zero automatically. With the double pixels per row technique and the zero padding, each SRAM bank of feature SRAM can transfer three pixels to the convolution core module.

\begin{figure}[h]
    \centering
    \includegraphics[width=8cm]{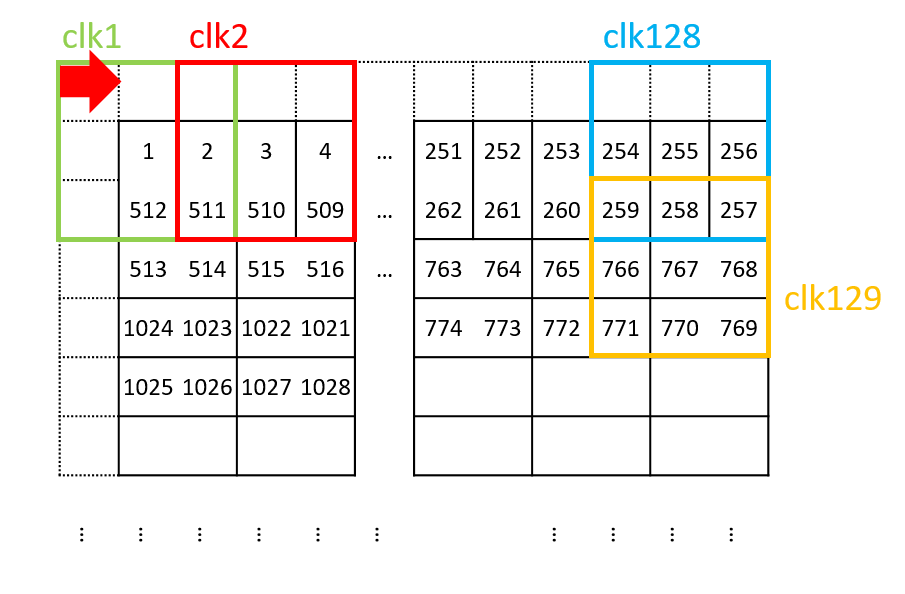}
    \caption{Data placement with DDPR in convolution with stride 2.}
    \label{fig:convolution_stride_2}
\end{figure}

The discussion above are the situation of the convolution with stride 1. However, the proposed double pixel per row technique can not only support the convolution with stride 1 but process the convolution with stride 2 without any negative effects. As shown in Fig. \ref{fig:convolution_stride_2}, from clk 1 to clk 2, four new data are needed. Since the DDPR and the dual-ports SRAM, our feature SRAM still can transfer four pixels in one cycle to fill the process element array in time at the front shifting. For the up shifting, feature SRAM will preload two rows of input feature map to ensure that six new data can transfer to process element array in the same cycle. For the left/right shifting, the preload data still can play a role to ensure the feature SRAM can provide four new data, which come from four different rows of feature SRAM.

\begin{figure}[h]
    \centering
    \includegraphics[width=8cm]{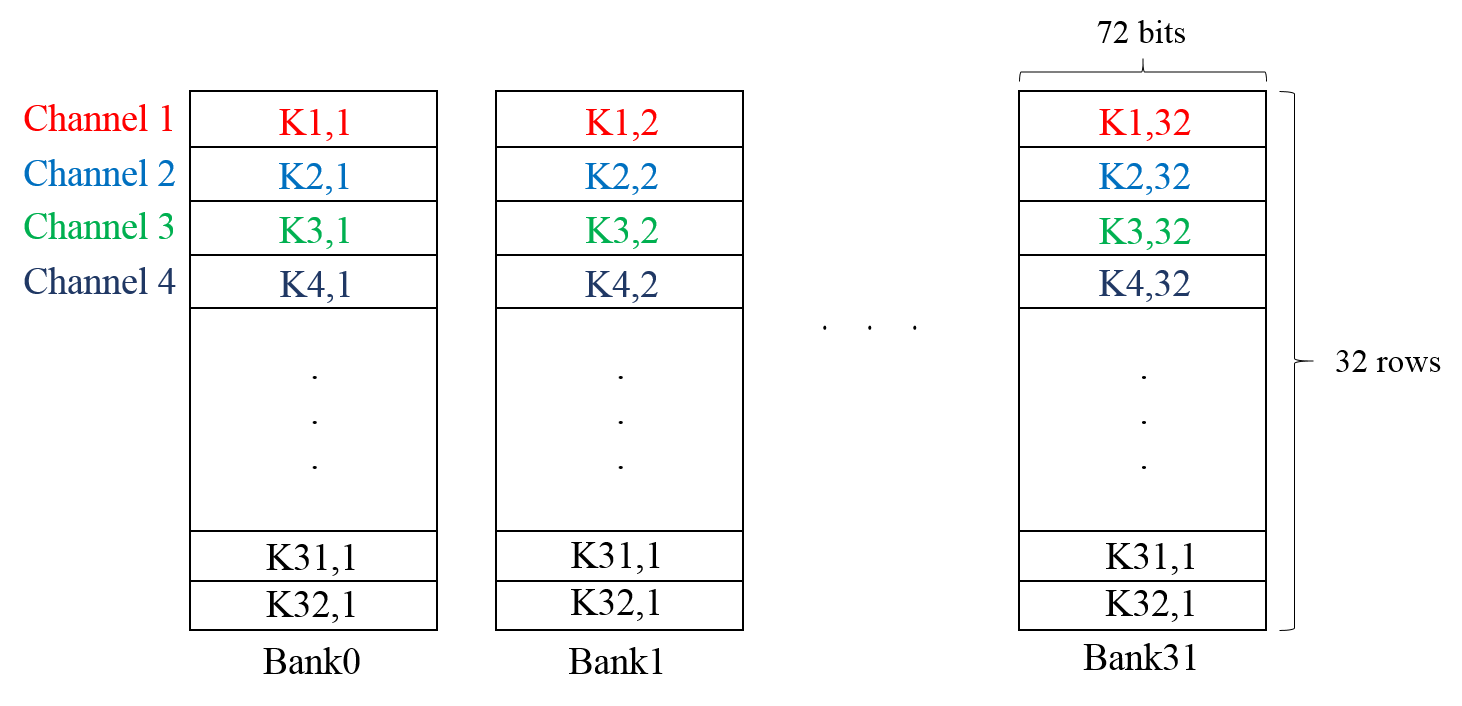}
    \caption{Data placement of proposed weight SRAM.}
    \label{fig:WSRAM}
\end{figure}

As shown in Fig.~\ref{fig:WSRAM}, we store kernel weights parallelly in 32 SRAM by channel. In order to reduce the initial time of transferring kernel weights from weight SRAM to convolution core module, each row of weight SRAM store nine weights, which can combined to a kernel. Since the proposed weight SRAM consist of 32 SRAM banks, it can make the time of kernel loading 32x faster.

In the end of this subsection, we introduce the last  part of our data unit, reuse SRAM. As shown in Fig.~\ref{fig:Data_unit}, data unit consists of two parts SRAM banks with different purposes. Here, we first introduce the feature reuse SRAM.

The purpose of feature reuse SRAM (FRSRAM) is to store the data of input feature map, which can be reuse. For tiling input feature map, we need to hold two rows in the end of the section of split input feature map. We store the input feature map by column in each row of feature reuse SRAM. Because of this data placement, we can transfers two pixels from feature reuse SRAM and one pixel from feature SRAM to process element array. Therefore, our proposed convolution core module still can calculate one output pixel in each cycle at the front shifting.

For the pooling reuse SRAM (PRSRAM), it is utilized to store the output feature map from the previous section. Each row of pooling reuse SRAM bank stores two output pixels. After the convolution of next section have a final sum in feature SRAM, pooling reuse SRAM will transfer two pixels and feature SRAM will transfer two corresponding pixels to the proposed pooling module to perform the 2x2 max pooling.

Table~\ref{tb:SRAM_size} shows the size of SRAM in each part of data unit. After sum up these different kinds of SRAMs with various sizes, we can get the total SRAM size of the accelerator, 289 KB.

\begin{table}[h]
\centering
\caption{SRAM size in each part.}
\label{tb:SRAM_size}
\setlength{\tabcolsep}{5mm}
\renewcommand{\arraystretch}{1.5}
\begin{tabular}{|c|c|}
\hline
               & Size (KB)    \\ \hline
Feature SRAM   & 256          \\ \hline
Weight SRAM    & 9            \\ \hline
Reuse SRAM     & 24           \\ \hline
Total & 289 \\ \hline
\end{tabular}
\end{table}

\subsubsection{Pooling Module}
\label{s:pooling_module}

Here, we introduce the proposed on-fly max pooling. If there has a max pooling layer after the current convolution layer, On-fly max pooling can do max pooling and convolution in the same time. Therefore, the we don't need to write back the whole original output feature map to the external DRAM for perform the max pooling. For the above reasons, the proposed on-fly max pooling can reduce the latency of inference and the data access amount of the external DRAM.

\begin{figure}[h]
    \centering
    \includegraphics[width=9cm]{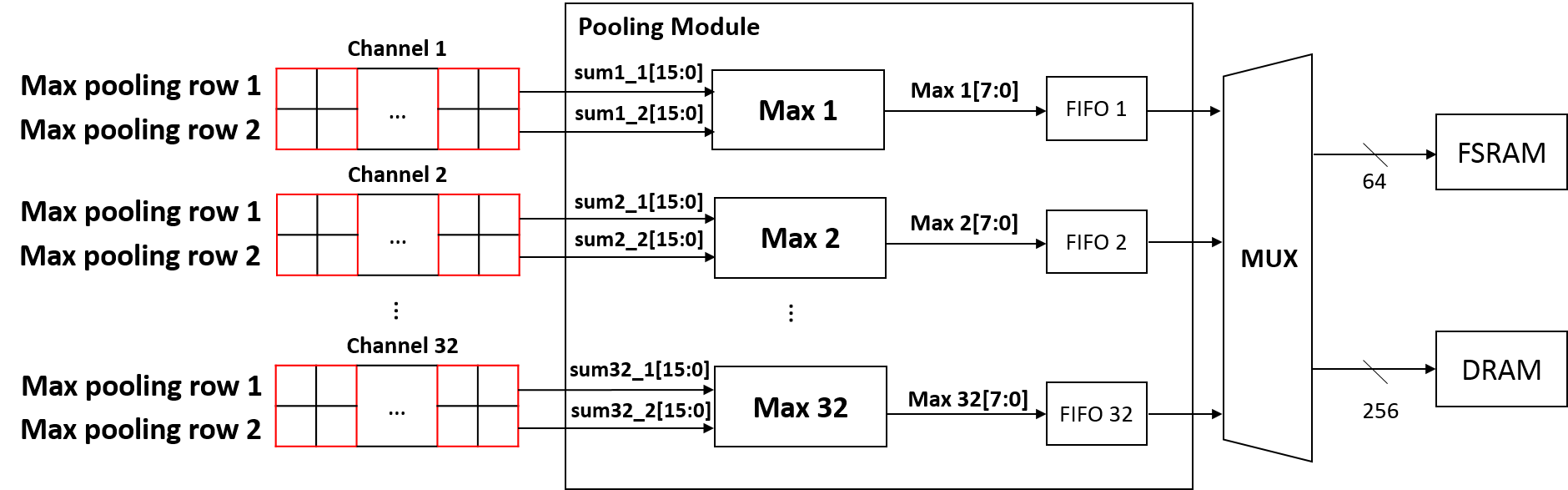}
    \caption{Overview of the proposed pooling module.}
    \label{fig:max_pooling_module}
\end{figure}

2x2 max pooling need four pixels from two rows of output feature map to execute as the red box. So, we need to wait for the calculate of the first row, after we can produce one max pooling result pixel every two cycles. Fig.~\ref{fig:max_pooling_module} shows the architecture of the proposed pooling module. Because of a dual-ports feature SRAM and the proposed double pixels per row data placement, four pixels is given once a cycle to do max-pooling. After the max operation, maximum values will be held in FIFO, which size is 128 $\times$ 8 bits, and the DEMUX will select where to output pixels. If the feature SRAM is big enough to store whole output feature map of current convolution layer, pooling module will store back the results to feature SRAM. Otherwise, pooling results will be stored back to the external DRAM. Even we still need to store the output feature map to DRAM, after the max pooling module the size of output feature map have been reduced by four times and save a lot of energy consumption.

\section{Experimental Results}
\label{sec:experimental_results}
In this chapter, we evaluate our proposed accelerator in three issues, energy efficiency, hardware utilization and computation performance. For the energy efficiency, we will discuss with the external DRAM access amount in section~\ref{s:external_DRAM_access_amount}. Then, discuss the hardware utilization in section~\ref{s:hardware_utilization} and computation performance in section~\ref{s:computation_performance}. We use three state-of-the-art models in our evaluation, RGBD eCNN~\cite{wang2020fast}, VGG-16~\cite{simonyan2014very} and MobileNet v1~\cite{howard2017mobilenets}. In section~\ref{s:synthesis_results}, we present the hardware implementation result of the proposed accelerator. Finally, we compare the implementation result of our proposed CNN accelerator with other works in section~\ref{s:comparison_with_other_works}.

\subsection{External DRAM Access Amount}
\label{s:external_DRAM_access_amount}

Table~\ref{tb:data_access_ecnn} shows the total amount of external DRAM access for both input and output reuse on RGBD eCNN with different proposed methods. The result shows the output reuse strategy can reduce about 252x of the DRAM access amount than the case, which do not use any reuse strategy. Moreover, the output reuse strategy only need the half data access amount of input reuse strategy. Therefore, our proposed accelerator choose to use the output reuse strategy. After we add the proposed on-fly pooling method, the reduce ratio of DRAM access amount can improve to 432x. At last, we further use the proposed ping-pong feature SRAM to reduce the DRAM access amount, and the total reduce about 533x with the no reuse case.

\begin{table}[h]
\centering
\caption{Data access amount comparison on RGBD eCNN model.}
\label{tb:data_access_ecnn}
\setlength{\tabcolsep}{2.5mm}
\renewcommand{\arraystretch}{1.5}
\begin{tabular}{|c|c|c|}
\hline
                          & DRAM Access Amount (MB) & Ratio \\ \hline
No Reuse                  & 1261.41                 & 1x    \\ \hline
Input Reuse               & 10.72                   & 117x  \\ \hline
Output Reuse              & 5                       & 252x  \\ \hline
Input Reuse + OFP         & 6.43                    & 196x  \\ \hline
Output Reuse + OFP        & 2.92                    & 432x  \\ \hline
Output Reuse + OFP + PPFS & 2.28                    & 533x  \\ \hline
\end{tabular}
\end{table}

For model VGG-16, we compare the DRAM access amount with the baseline paper, \cite{chang2019vwa}'s approach. Because of the data precision in \cite{chang2019vwa} is 16 bits, we halve the DRAM access amount of each convolution layer to make the comparison equitable. As table~\ref{tb:data_access_VGG_16} shows, our output reuse strategy can reduce more DRAM data access amount than \cite{chang2019vwa}. And the proposed on-fly pooling and ping-pong feature SRAM can further reduce the DRAM data access amount to about 72.33 MB. Overall our methods can reduce about 252x of the DRAM access amount than the \cite{chang2019vwa}'s approach.

\begin{table*}[ht]
\centering
\caption{Comparison of DRAM access amount between \cite{chang2019vwa}'s approach and ours on VGG-16}
\label{tb:data_access_VGG_16}
\setlength{\tabcolsep}{5mm}
\renewcommand{\arraystretch}{1.5}
\begin{tabular}{|c|c|c|c|c|c|}
\hline
 & \cite{chang2019vwa} & \cite{chang2019vwa} & Output Reuse & \begin{tabular}[c]{@{}c@{}}Output Reuse\\ + OFP\end{tabular} & \begin{tabular}[c]{@{}c@{}}Output Reuse\\ + OFP + PPFS\end{tabular} \\ \hline
Data Precision & fixed 16 bits & fixed 8 bits & fixed 8 bits & fixed 8 bits & fixed 8 bits \\ \hline
Total (MB) & 202.968 & 101.484 & 90.295860 & 78.620079 & 72.332971 \\ \hline
Ratio & - & 1x & 1.12x & 1.29x & 1.4x \\ \hline
\end{tabular}
\end{table*}

For model MobileNet v1, we compare the DRAM access amount with the our methods and case with no any reuse strategy. Since MobileNet does not use the max pooling layer to reduce the size of input feature map, so the proposed on-fly pooling method will not work on this model. As shown in table~\ref{tb:data_access_mobilenet}, we use the output reuse strategy and the ping-pong feature SRAM to reduce the DRAM access amount.

\begin{table}[H]
\centering
\caption{Data access amount comparison on MobileNet v1 model.}
\label{tb:data_access_mobilenet}
\setlength{\tabcolsep}{3mm}
\renewcommand{\arraystretch}{1.5}
\begin{tabular}{|c|c|c|}
\hline
                    & DRAM Access Amount (MB) & Ratio \\ \hline
No Reuse            & 2044.77                 & 1x    \\ \hline
Output Reuse        & 58.3                    & 35x   \\ \hline
Output Reuse + PPFS & 23.98                   & 86x   \\ \hline
\end{tabular}
\end{table}

\subsection{Hardware Utilization}
\label{s:hardware_utilization}

Because of the model structure of RGBD eCNN, we can maintain 100\% utilization in whole convolutional time. Since the full utilization of hardware resource, our computation performance can reach to the peak throughput of the proposed accelerator.

\begin{figure*}[h]
\begin{center}
\subfigure[Hardware utilization of the proposed accelerator on VGG-16.] {
	\includegraphics[width=8cm]{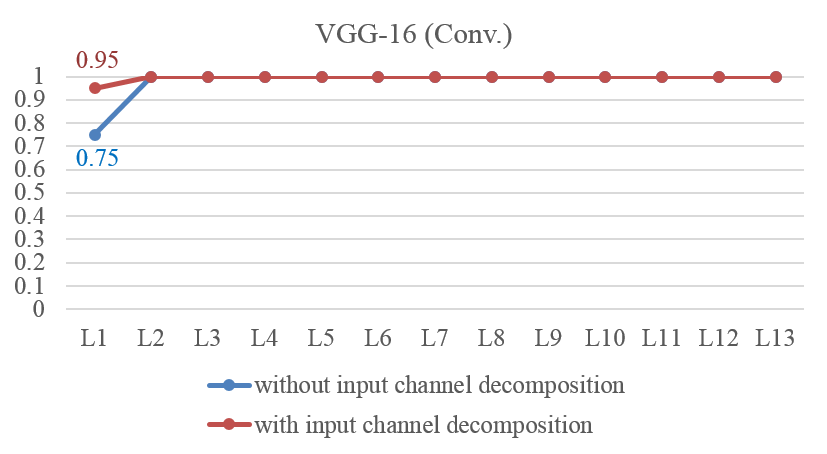}
	\label{fig:utilization_VGG16_ours}
}
\subfigure[Hardware utilization \cite{chang2019vwa}'s approach on VGG-16.] {
	\includegraphics[width=8cm]{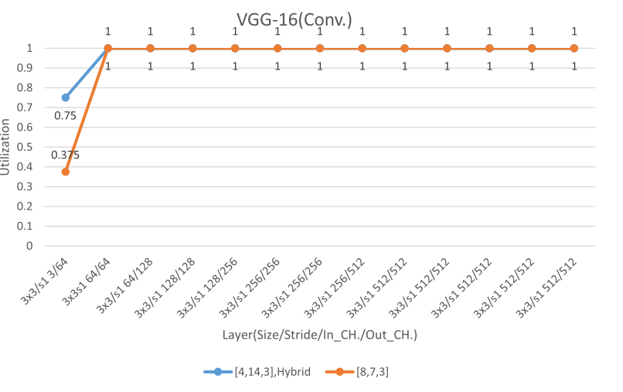}
	\label{fig:Utilization_VGG16_[2]}
}
\caption{Hardware utilization of (a) Our proposed accelerator and (b) \cite{chang2019vwa}'s approach on VGG-16.}
\label{fig:utilization_VGG16}
\end{center}
\end{figure*}

For model VGG-16, we compare the hardware utilization with the baseline paper and \cite{chang2019vwa}'s approach. Because of the proposed input channel decomposition, we can improve the hardware utilization of layer 1 from 75\% to 95\% as Fig.~\ref{fig:utilization_VGG16_ours}. As shown in Fig.~\ref{fig:utilization_VGG16}, both of accelerator can maintain full hardware utilization in layer 2 to layer 13, but in layer 1 we can improve 20\% utilization than \cite{chang2019vwa}'s approach.

For MobileNet v1, depthwise convolution consist of separable convolution and pointwise convolution. Pointwise convolution is a normal convolution with 1x1 kernel, but the separable convolution only need to process each channel of input feature map with one 3x3 kernel. Therefore, separable convolution only can use one process element array in a row of convolution core module.

The hardware utilization of separable convolution is only 25\%, and the hardware utilization of pointwise convolution can reach 89\%. However, the computation of separable convolution only accounted for 30\% in whole depthwise convolution, so the hardware utilization of depthwise convolution still can reach 70\% as MobileNet layer 2 for example.

\subsection{Computation Performance}
\label{s:computation_performance}

\quad Since our high utilization in RGBD eCNN, the proposed accelerator can reach the peak throughput in real condition. Table~\ref{tb:compare_latency_ECNN} shows the comparison of inference latency on RGBD eCNN with CPU, GPU, and two kinds RGBD eCNN accelerator, eCNN~\cite{huang2019ecnn} and eCNN Lite~\cite{wang2020fast}. Compare the inference latency with the eCNN Lite CNN accelerator, the proposed accelerator can speed up about 2.57x.

\begin{table*}[h]
\centering
\caption{Compare latency with the other works on RGBD eCNN model.}
\label{tb:compare_latency_ECNN}
\setlength{\tabcolsep}{3mm}
\renewcommand{\arraystretch}{1.5}
\begin{tabular}{|c|c|c|c|c|c|}
\hline
Hardware Platform & Intel i7            & GTX 1080         & eCNN Lite~\cite{wang2020fast} & eCNN~\cite{huang2019ecnn} & Ours     \\ \hline
Inference Latency & 101.5$\sim$125.1 ms & 1.3$\sim$14.3 ms & 4.506 ms  & 0.171 ms & 1.752 ms \\ \hline
\# of MAC          & -       & -     & - & 81920 & 1152 \\ \hline       
Speed Up          & -                   & -                & 1x       & -        & 2.57x    \\ \hline
\end{tabular}
\end{table*}

As shown in table~\ref{tb:compare_latency_VGG16}, Our proposed accelerator has 1152 PE number, 6.8x more than \cite{chang2019vwa} and \cite{chen2016eyeriss}. However, ours accelerator can has 7.54x speed up than \cite{chang2019vwa} that is because the proposed input channel decomposition can improve the hardware utilization in layer 1 to reduce the inference latency. In addition, the inference latency of the proposed accelerator is 13.35 ms means we can inference VGG-16 with 75 FPS. In other words, the proposed accelerator can real-time inference VGG-16.

\begin{table}[H]
\centering
\caption{Compare latency with other works on VGG16.}
\label{tb:compare_latency_VGG16}
\setlength{\tabcolsep}{3mm}
\renewcommand{\arraystretch}{1.5}
\begin{tabular}{|c|c|c|c|}
\hline
                & Ours      & \cite{chang2019vwa}  & \cite{chen2016eyeriss}   \\ \hline
PE number       & 1152      & 168      & 168       \\ \hline
Clock rate      & 500 Mhz   & 500 Mhz  & 200 Mhz   \\ \hline
Peak throughput & 1152 Gops & 168 Gops & 67.2 Gops \\ \hline
Total Latency (ms)   & 13.35139  & 100.69   & 1436.47   \\ \hline
Speed Up        & 7.54x   & 1x   & -        \\ \hline
\end{tabular}
\end{table}

For MobileNet v1, since the high utilization of 1x1 convolution, the inference latency is 2.13 ms. Table~\ref{tb:compare_latency_MobileNet} shows the comparison of FPS with \cite{zhao2018automatic}, the proposed accelerator can speed up about 2x than \cite{zhao2018automatic}'s approach.

\begin{table}[H]
\centering
\caption{Compare the inference latency with other work on MobileNet v1.} \label{tb:compare_latency_MobileNet}
\setlength{\tabcolsep}{3mm}
\renewcommand{\arraystretch}{1.5}
\begin{tabular}{|c|c|c|}
\hline
                 & \cite{zhao2018automatic} & Ours   \\ \hline
Inference Latency (ms) & 4.32 & 2.13 \\ \hline
Frame per Second & 231.7   & 468.76 \\ \hline
Speed Up         & 1x      & 2x     \\ \hline
\end{tabular}
\end{table}

\subsection{Synthesis Results}
\label{s:synthesis_results}

\quad We implement the proposed accelerator by Verilog and synthesized at 500 Mhz with DC compiler under TSMC 40 nm technology. Table~\ref{tb:synthesis_result_proposed} lists the synthesized results of the proposed accelerator, operate at 500 Mhz while area size is about 3.59 $mm^2$ and power comsumption is about 554.57 mW. Within the computation performance, energy efficiency of the proposed accelerator can reach 2.08 Tops/W. 

\begin{table}[H]
\centering
\caption{Synthesis results of the proposed accelerator.} \label{tb:synthesis_result_proposed}
\setlength{\tabcolsep}{3mm}
\renewcommand{\arraystretch}{1.5}
\begin{tabular}{|c|c|c|}
\hline
\multicolumn{3}{|c|}{Proposed Accelerator}                            \\ \hline
Area ($mm^2$) & Power (mW)  & Energy Efficiency (Tops/W) \\ \hline
3.59   & 554.5682               & 2.08                       \\ \hline
\end{tabular}
\end{table}

\subsection{Comparison with other works}
\label{s:comparison_with_other_works}

\begin{table*}[h]
\centering
\caption{Comparison with other works.} 
\label{tb:synthesis_result_proposed}
\setlength{\tabcolsep}{3mm}
\renewcommand{\arraystretch}{1.5}
\begin{tabular}{|c|c|c|c|c|c|c|}
\hline
                              & Our Work & \cite{huang2019ecnn} & {\cite{chang2019vwa}}  & \cite{chen2016eyeriss} & \cite{du2017reconfigurable} & \cite{tu2017deep}  \\ \hline
Technology                    & 40 nm    & 40 nm   & 40 nm    & 65 nm    & 65 nm    & 65 nm    \\ \hline
Clock Rate (Mhz)              & 500      & 250     & 500      & 250      & 500      & 200      \\ \hline
\#   of MAC                   & 1152     & 81920   & 168      & 168      & 168      & 512      \\ \hline
Performance   (Gops)          & 1152     & 41000   & 168      & 42       & 152      & 204.8    \\ \hline
Area ($mm^2$)                       & 3.59     & 55.23   & 1.56     & 12.25    & 5        & 16       \\ \hline
SRAM   (KB)                   & 289      & 2864    & 191      & 181.5    & 96       & 280      \\ \hline
Power   (mW)                  & 554.57   & 6940    & 154.98   & 278      & 354      & 479      \\ \hline
Bit-Width   (bits)            & 8 fixed  & 8 fixed & 16 fixed & 16 fixed & 16 fixed & 16 fixed \\ \hline
Power   Efficiency   (Tops/W) & 2.08     & 5.9     & 1.084    & 0.151    & 0.429    & 0.406    \\ \hline
\end{tabular}
\end{table*}

Table~\ref{tb:synthesis_result_proposed} shows the comparison with other works. The purposed accelerator have a good performance compares to the other works, only lower than \cite{huang2019ecnn}, but \cite{huang2019ecnn} uses too much hardware resource that make the it has the huge area size. Compared to \cite{chang2019vwa}, we not only have the better performance but have great power efficiency.

\section{Conclusion}
\label{chap:Conclusion}

In this thesis, we propose a high utilization energy-aware real-time inference DCNN accelerator design to tackle three important issues, hardware utilization, external memory access and computation complexity.

The hardware implementation of the proposed accelerator architecture under the TSMC 40 nm technology reaches 1.152 Tops/s with 554.57 mW total power in 3.59 $mm^2$ area size.

Compared the proposed CNN accelerator to \cite{chang2019vwa} on VGG16, we have higher hardware utilization in convolution layer 1 to speed up the inference time about 7.5x and reduce the data access amount of external DRAM about 1.4x, which is the most energy consumption part for the accelerator inference the CNN model. Therefore, the energy efficiency of the proposed accelerator reach 2.08 Tops/W, which is 1.92x higher than \cite{chang2019vwa}.

\bibliographystyle{IEEEtran}
\bibliography{reference}

\end{document}